\documentclass[twocolumn]{aastex62}

\usepackage{natbib}
\usepackage{color}
\usepackage{graphicx}
\usepackage[caption=false]{subfig}
\usepackage{mathtools}
\graphicspath{{./}{figures/}}

\submitjournal{ApJ}

\shorttitle{Size Evolution}
\shortauthors{A. Whitney et al.}

\begin{document}

\title{Unbiased Differential Size Evolution and the Inside-Out Growth of Galaxies in the \\ Deep CANDELS GOODS Fields at $1 \leq z \leq 7$}

\author{A. Whitney}
\affil{University of Nottingham, School of Physics \& Astronomy, Nottingham, NG7 2RD, UK}

\author{C. J. Conselice}
\affiliation{University of Nottingham, School of Physics \& Astronomy, Nottingham, NG7 2RD, UK}

\author{R. Bhatawdekar}
\affiliation{University of Nottingham, School of Physics \& Astronomy, Nottingham, NG7 2RD, UK}

\author{K. Duncan}
\affiliation{Leiden Observatory, Leiden University, PO Box 9513, NL-2300 RA Leiden, The Netherlands}



\begin{abstract}

We present a size analysis of a sample of $\sim$ 49,000 galaxies from the CANDELS GOODS North and South fields using redshift-independent relative surface brightness metrics to determine an unbiased measure of the differential size evolution of galaxies at $1 \leq z \leq 7$. We introduce a novel method of removing foreground objects from distant galaxy ($z > 3$) images that makes use of the Lyman-break at 912{\AA}, in what we call `2-D Lyman-Break Imaging'. The images used are in the rest-frame optical at $z < 3$ and progressively bluer bands at $z > 3$. They are therefore subject to K-correction and cosmological dimming effects which are tested and corrected for. We separately consider a mass-selected sample (with masses in the range 10$^9$M$_{\odot}$$\leq$M$_*$$\leq$10$^{10.5}$M$_{\odot}$) and a number density selected sample (using a constant number density of $n = 1\times10^{-4}$Mpc$^{-3}$). Instead of utilising the commonly used, but potentially biased, effective radii for size measurements, we measure the redshift-independent Petrosian radius, defined by the parameter $\eta$, for each galaxy for three values of $\eta$ and use this as a proxy for size. The evolution of the measured radii can be described by a power-law of the form $R_{\textup{Petr}} = \alpha(1+z)^\beta$kpc where $\beta < 0$. We find that the outer radius increases more rapidly, suggesting that as a galaxy grows mass is added to its outer regions via an inside-out growth. This growth is stronger for the number density selected sample, with a growth rate of nearly three in the outer radii compared to the inner. We test and confirm these results using a series of image simulations.


\vspace{10mm}

\end{abstract}

\keywords{}


\section{Introduction} \label{sec:intro}

Despite extensive research, the details of the processes that form and influence a galaxy's evolution are still largely unknown. This will change over the next decade with the advent of new facilities such as the \textit{James Webb Space Telescope}. However, there is still a great deal of information to be gathered from existing data from the \textit{Hubble Space Telescope}. 

The size evolution of galaxies through redshift can tell us critical information about the potential formation scenarios undergone by the first galaxies in order to produce the galaxies seen at later times. The size of a galaxy is one of the easiest and most direct properties that can be measured. The effective radius has typically been used to determine this aspect of galaxy evolution over a range of redshifts e.g. \cite{buitrago08, wel08, bouwens04, allen17}. 

It has been shown using the \textit{Hubble Space Telescope} Advanced Camera for Surveys (ACS) and GOODS Near Infrared Camera and Multi-Object Spectrometer (NICMOS) \citep{conselice11} data that there is a strong evolution in the effective radii of galaxies since $z = 3$ \citep{trujillo07, buitrago08, cassata13}. This is further confirmed by the use of data from the Cosmic Assembly Near-infrared Deep Extragalactic Survey (CANDELS) \citep{grogin11,koekemoer11} and NICMOS data \citep{dokkum08, buitrago08, weinzirl11, bruce12, buitrago13, lani13, patel13, wel14}. Ground based observations also yield similar results \citep{dokkum10, carrasco10}. These studies show a size evolution with an increase in effective radius since $z \sim 2$ from a factor of $\sim$2 \citep{wel08} up to a factor of $\sim$7 \citep{buitrago08, carrasco10}. This strong evolution in size is consistent with simulations of massive galaxies forming through minor mergers \citep{naab09, furlong17}.

At redshifts higher than $z = 3$, a less steep evolution in size is found compared to $z < 3$ with the effective radius changing as $(1 + z)^{-\beta}$ where $\beta$ $\simeq$ 1 up to $z = 7$ \citep{bouwens04, oesch10, straatman15, curtislake16, allen17}. Probing higher up to a redshift of $z \simeq 12$, the measured sizes also fit with extrapolated data \citep{ono13, holwerda15}. Such studies, both at low and high redshifts, show that massive galaxies at $z > 1$ are significantly more compact than galaxies of a comparable mass at low redshift \citep{ferguson04, cimatti08, damjanov09}. Observations of galaxies at high redshifts are subject to cosmological dimming and K-correction effects and this can lead to finding no obvious evolution in size with redshift \cite[e.g.][]{law07, ichikawa12, ribeiro16}.

The cause of the observed evolution in size is thought to be a result of accretion of gas and stars from the intergalactic medium and mergers with other galaxies \citep[e.g.][]{ferreras09, lopezsanjuan12}. \cite{conselice13} and \cite{ownsworth16} find that accretion is the dominant formation mode amongst the most massive galaxies by calculating the evolution of stellar mass from observed star formation rates and the amount of stellar material added via mergers. \cite{bluck12} suggests mergers are the primary cause of the observed size evolution in massive galaxies by determining the merger history of a sample of galaxies from the GOODS NICMOS Survey. Additionally, \cite{bluck12} suggest mergers can explain the majority of size evolution since $z \sim 1$ assuming mergers occur over a short timescale. \cite{naab09} also show this by simulating the formation of a massive spheroidal galaxy. The cosmological hydrodynamical simulation Illustris has also been used to show that the growth in size experienced by galaxies is largely caused by mergers \citep[e.g.][]{wellons16}. It has also been suggested that this evolution is due to quasar feedback that removes gas from central regions, which in turn induces the expansion of the stellar distribution \citep{fan08}. However, the details of the processes that lead to the growth of galaxies over time are still largely unknown. 

Through the study of the evolution of sizes in a sample of galaxies, we can hope to expand the understanding of the formation of galaxies. However a more refined method is now needed to make further progress on the study of galaxy sizes, and those measured using parametric fitting are often subject to biases produced by redshift. Thus, in this paper we make use of the Petrosian radius, a redshift-independent measure of the size determined by the ratio of surface brightness at a particular radius and the surface brightness within that radius \citep[e.g.][]{bershady00, conselice03b}. It allows us to determine not only whether galaxies are growing but also where the size is growing within the galaxies, i.e. whether the inner or outer regions of a galaxy are getting larger. By using this redshift-independent measure of size in combination with a new method to remove field galaxies from images, we present in this paper an unbiased view of how galaxies are changing in size over time. 

The structure of the paper is as follows: In Section \ref{sec:data}, we describe the data and the sample used. In Section \ref{sec:method} we describe the methods used to remove field objects from the postage stamp images of the galaxies in our sample and to calculate the sizes of the galaxies. In Section \ref{sec:results} we present our results. Finally, in Section \ref{sec:summ} we discuss our results, and their implications and present our conclusions in Section \ref{sec:conc}. Throughout this paper we use AB magnitudes and assume a $\Lambda$CDM cosmology with H$_0$ = 70 kms$^{-1}$Mpc$^{-1}$, $\Omega_m$ = 0.3, and $\Omega_{\Lambda}$ = 0.7.

\section{Data and Sample Selection} \label{sec:data}

The data we use in this paper are taken from the Advanced Camera for Surveys (ACS) and the Wide Field Camera 3 (WFC3) of the \textit{Hubble Space Telescope} (\textit{HST}). The fields used are the GOODS North and South fields of the Cosmic Assembly Near-infrared Deep Extragalactic Survey (CANDELS) \citep{grogin11, koekemoer11}. CANDELS covers a total area of 800 arcmin$^2$ over 5 different fields. GOODS North and GOODS South each cover an area of 160 arcmin$^2$ and are centered on the Hubble Deep Field North and the Chandra Deep Field South respectively \citep{giavalisco04}. Both GOODS fields were part of the Deep and Wide tiers of CANDELS which were observed using the WFC3 on \textit{HST} \citep{grogin11, koekemoer11} and these regions were observed in the F105W (Y$_{105}$), F125W (J$_{125}$), and F160W (H$_{160}$) filters. The ACS was used to observe the two fields in the F435W (B$_{435}$), F606W (V$_{606}$), F775W (i$_{775}$), F814W (I$_{814}$), and F850LP (z$_{850}$) filters. We use only the Deep tier and this covers the central regions of the GOODS fields.

\begin{figure}[!ht]
\includegraphics[width=0.475\textwidth]{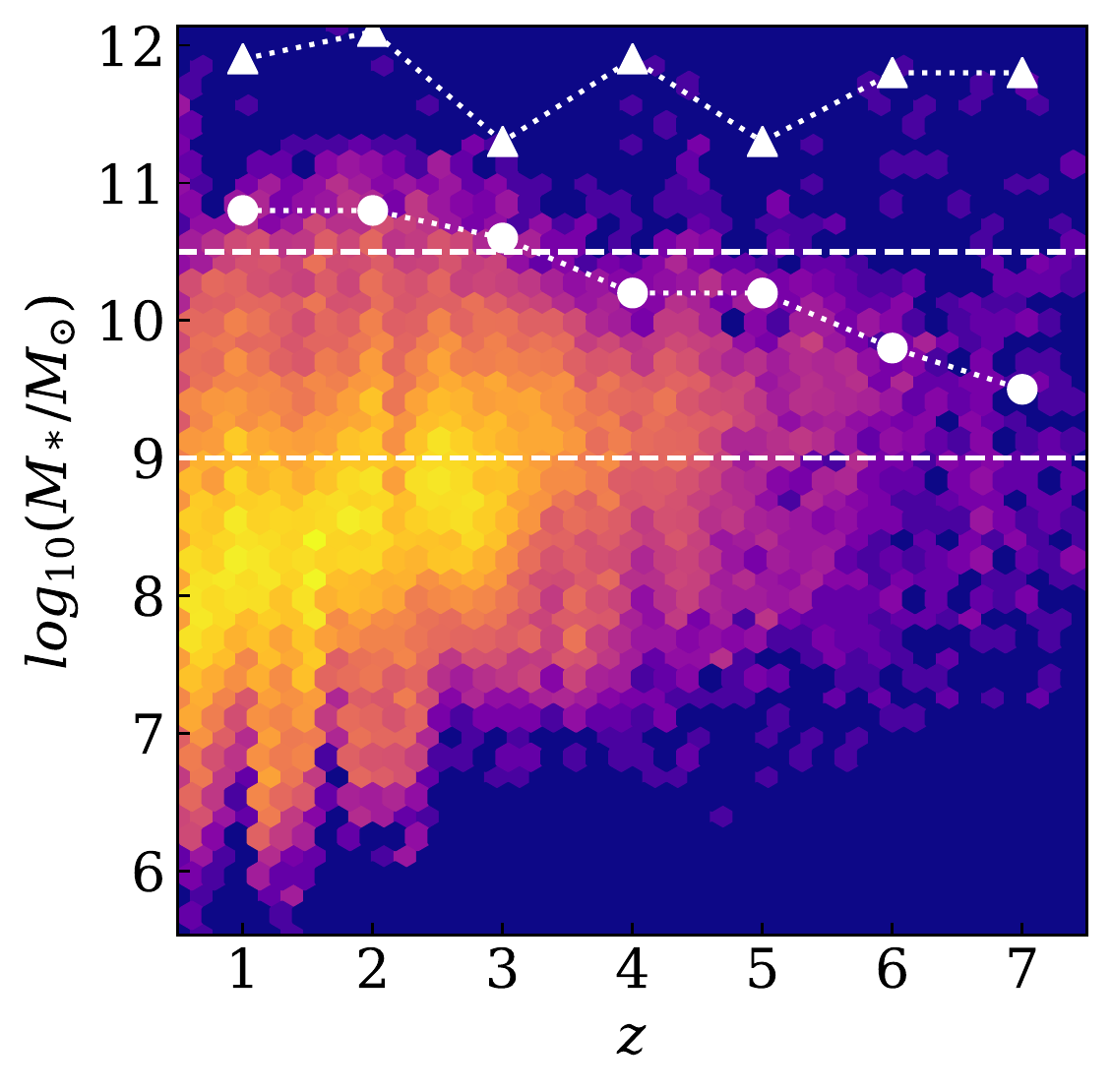}
\caption{The galaxy stellar mass-redshift distribution for all 48,575 galaxies in our sample from the entire CANDELS area of the GOODS North and South fields (see \cite{duncan14, duncan19} for details). The colours show the density of points with yellow representing the highest density and dark purple representing the lowest. The completeness limits are log$_{10}(M_{\odot})$ = 8.55, 8.685, 8.85, 9.15 for $z \sim$ 4, 5, 6, and 7 respectively \citep{duncan14}. The limits for $z < 3$  are considerably lower than our chosen lower mass limit of $10^{9}$M$_{\odot}$. The mass limits for each of the samples described in \ref{sec:results} are shown in white; the limits of the mass selected sample are shown as horizontal dashed lines. The upper mass limits for each of the number density selected sample are shown as triangles and the lower mass limits are shown as circles.}
\label{fig:Mzd}
\end{figure}

Our sample consists of 48,575 galaxies from both GOODS North and South fields, covering a redshift range of $1 \leq z \leq 7$ and a mass range of $10^6M_{\odot} \leq M_* \leq 10^{12}M_{\odot}$. Details of our sample are described in \cite{duncan14, duncan19} who make new estimates for the galaxy stellar mass function and star formation rates for this sample of galaxies in the CANDELS fields. Figure \ref{fig:Mzd} shows the distribution of the stellar mass and the redshift of the galaxies in our sample with the yellow bins representing the highest density of points and the dark purple representing the lowest density. We see that the the highest concentration of mass and redshift lies at approximately $10^8M_{\odot} \leq M_* \leq 10^{9}M_{\odot}$ and $1 \leq z \leq 3$ however the sample spans all redshifts and masses. When selecting galaxies at high redshift ($z > 6$) for our sample, we visually inspect the images to remove any contaminating galaxies that are potential false positive detections. To determine which objects were false positives, any that had no object visible in the H$_{160}$ band image or were saturated in H$_{160}$ band image were removed from the sample. Figure \ref{fig:Mzd} also shows the mass limits of each of the samples described in Section \ref{sec:results}. The mass limits of the mass selected sample are shown as horizontal dashed lines at log$_{10}(M_*/M_{\odot})$ = 9 and log$_{10}(M_*/M_{\odot})$ = 10.5. The upper mass limits of each redshift bin of the number density selected sample (constant number density of 1$\times$10$^{-4}$ Mpc$^{-3}$) are shown as triangles and the lower limits are shown as circles.

\subsection{Photometric Redshifts}

The photometric redshifts we use are calculated with the \textsc{eazy} photometric redshift software \citep{brammer08} by fitting all available \textit{HST} bands to a template based on the PEGASE spectral models of \cite{fioc97}. An additional very blue template based on a spectrum by \cite{erb10} is also used, and it includes features expected in young galaxy populations such as high Lyman-$\alpha$ equivalent widths and strong optical lines. The full redshift probability distribution function (PDF) is constructed for each galaxy by using the $\chi^2$-distribution produced by \textsc{eazy}. No magnitude based prior was included in the fitting due to large uncertainties in the H-band luminosity function at higher redshifts \citep{henriques12}.

Where available, the calculated photometric redshifts are compared to spectroscopic redshifts and there is a small scatter of $\sigma_{z,O} = \textup{rms}(\Delta z/(1 + z_{spec}))$ = 0.037 where $\Delta z = (z_{spec} - z_{phot})$. There is also a very small bias in the values with a median value of $\Delta z$ = -0.04. As such, the calculated photometric redshifts are similar to the spectroscopic redshifts from other sources.

To ensure the accuracy of the calculated photometric redshifts, the redshift for each galaxy is randomly taken from its PDF 500 times and the results are then averaged. For further details of the process, see \cite{duncan14} and \cite{duncan19}.

\subsection{Mass Fitting}

The stellar masses used here are determined by using the custom template fitting code \textbf{SMpy}\footnote{https://github.com/dunkenj/smpy} \citep{duncan14} with spectral energy distributions (SEDs) from the synthetic stellar population models of \cite{bruzual03}. Emission lines and continuum are added to the templates in line with previous high-redshift fitting methods, e.g. \cite{ono10,schaerer10,mclure11,salmon15}. For full details on the mass fitting process, see Section 4 of \cite{duncan14}.

\section{Methodology} \label{sec:method}

\subsection{2-D Lyman-Break Imaging}


Here we describe the new method we use to produce images to measure the properties of galaxies in our sample. The method uses the well-known Lyman-break drop technique where a galaxy at a certain redshift `disappears' at wavelengths redder than the Lyman-limit at 912{\AA} which creates a sharp break in the continuum. This break gives galaxies distinctive UV rest-frame colours which can be used to select galaxies at specific redshifts using photometry within multiple filters. The hydrogen gas absorbs the bluest wavelengths of light and thus the target object essentially disappears or becomes significantly fainter compared to the flux in redder bands. The break is observed at redder wavelengths as the redshift increases. For galaxies at $z = 6$ and $z = 7$, the break falls within the V$_{606}$ band, and for galaxies at $z = 5$ and $z = 4$, the break falls within the B$_{435}$ band. The band corresponding to the break for galaxies at $z < 3$ is at a shorter wavelength than the available filters therefore this image processing technique is applied only to galaxies at $z \geq 4$. Note that effectively the galaxy nearly disappears in filters which probe light below the Lyman limit, thus making it possible to isolate the light which belongs to these galaxies from those at lower and significantly higher redshifts. However, the majority of field objects will be at a lower redshift than the target objects so there will be minimal contamination from those that are at higher redshifts.

\begin{figure*}
\centering
\begin{tabular}{cccc}
\LARGE{$D_{i,j}^{raw}$} & \LARGE{$O_{i,j}^{raw}$} & \LARGE{$S_{i,j}$} & \LARGE{$O_{i,j}^{analysis}$} \\
\subfloat{\includegraphics[width = .23\textwidth]{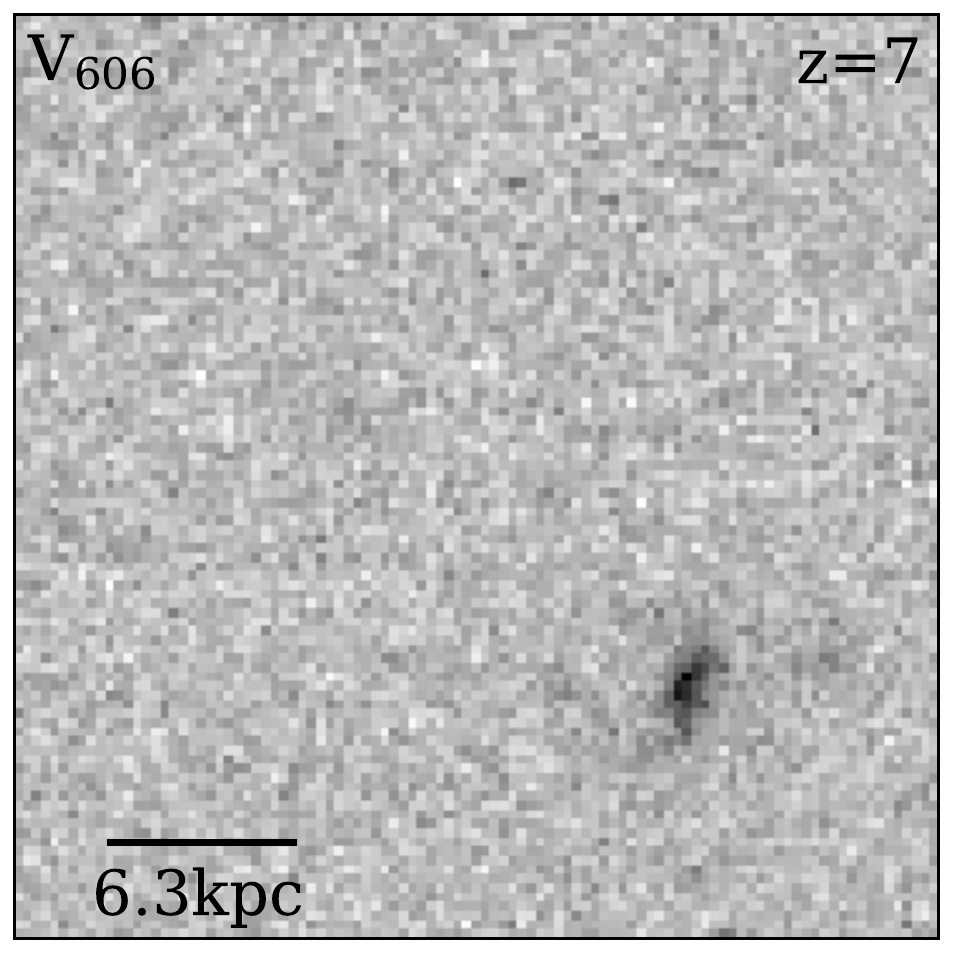}} &
\subfloat{\includegraphics[width = .23\textwidth]{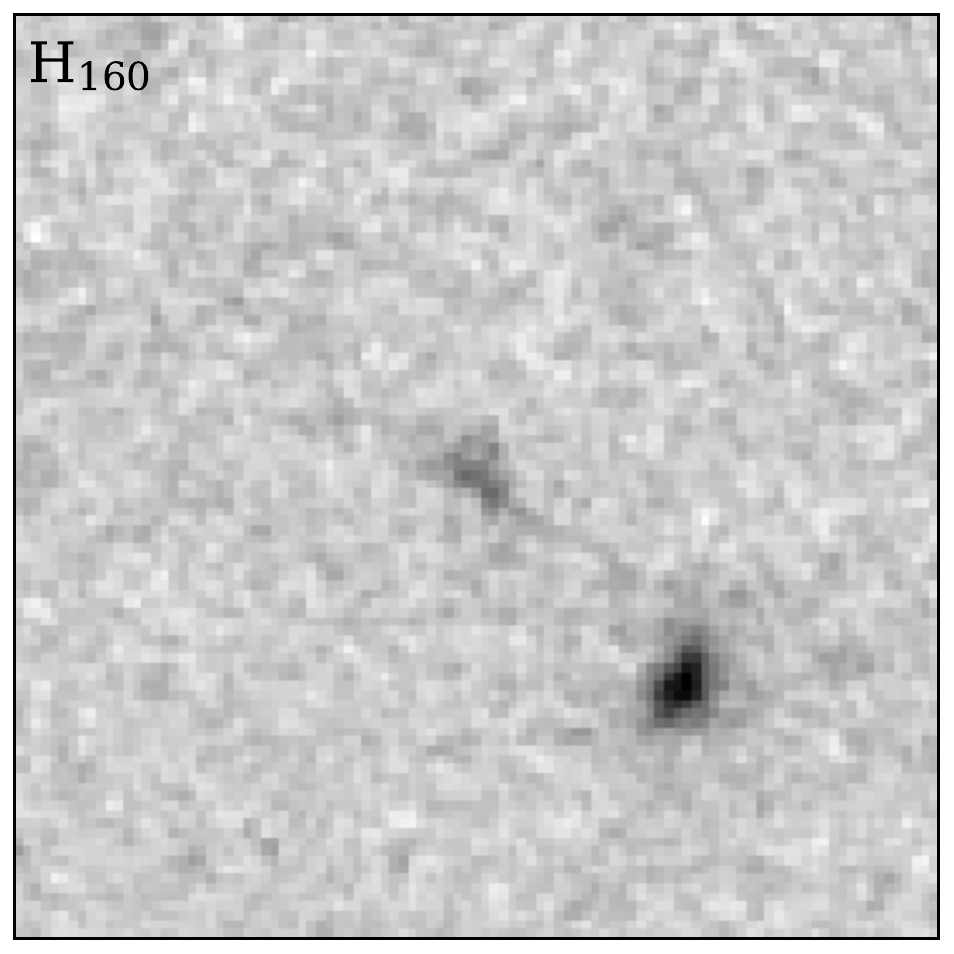}} &
\subfloat{\includegraphics[width = .23\textwidth]{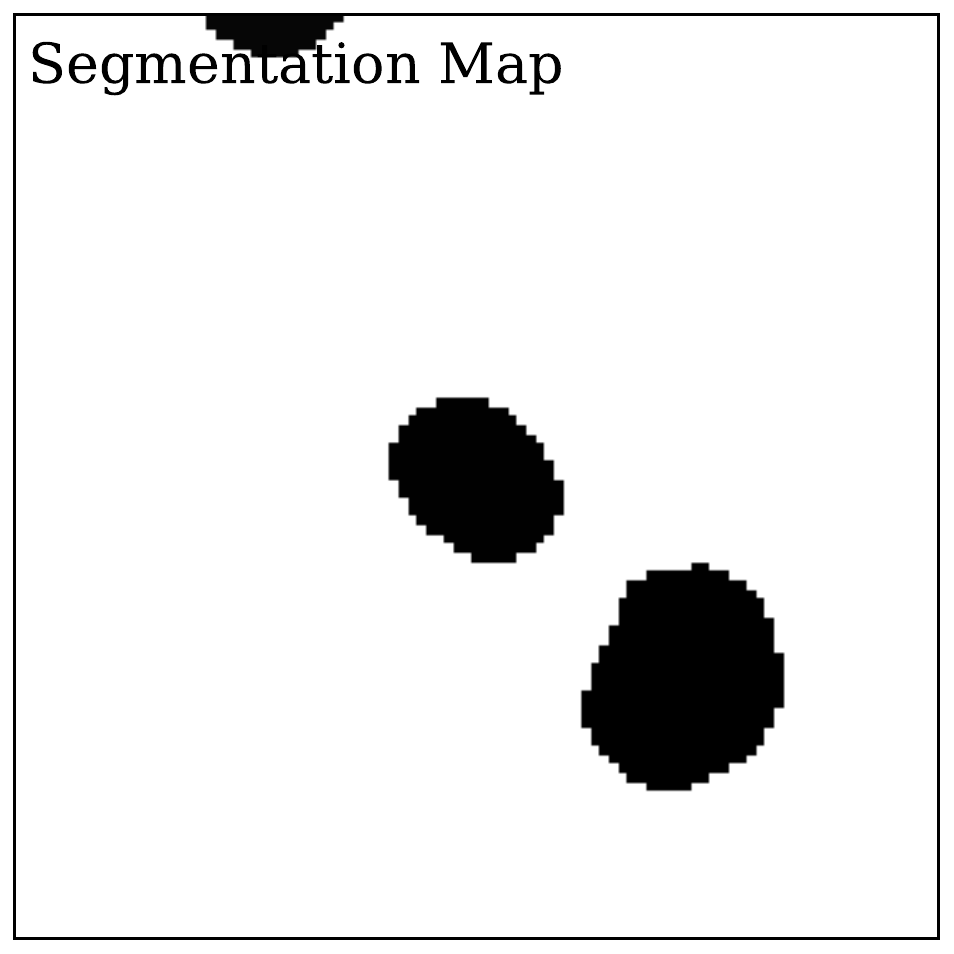}} &
\subfloat{\includegraphics[width = .23\textwidth]{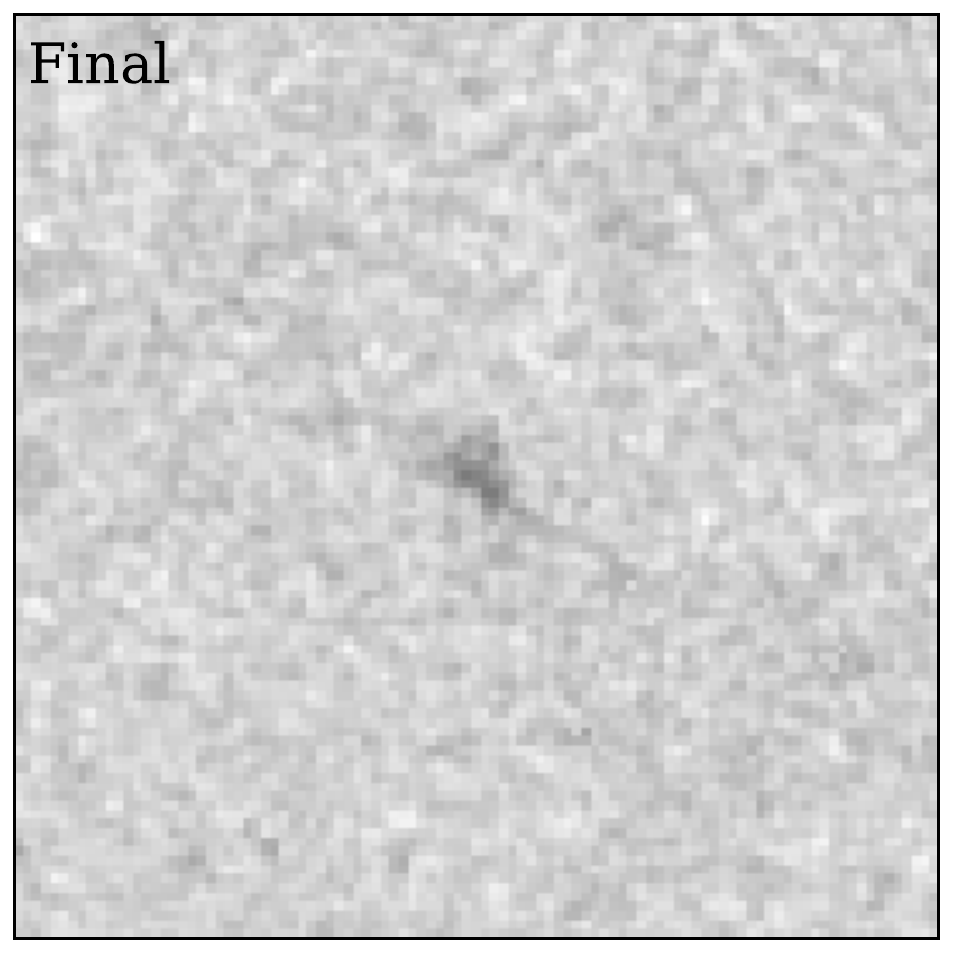}} \\
\subfloat{\includegraphics[width = .23\textwidth]{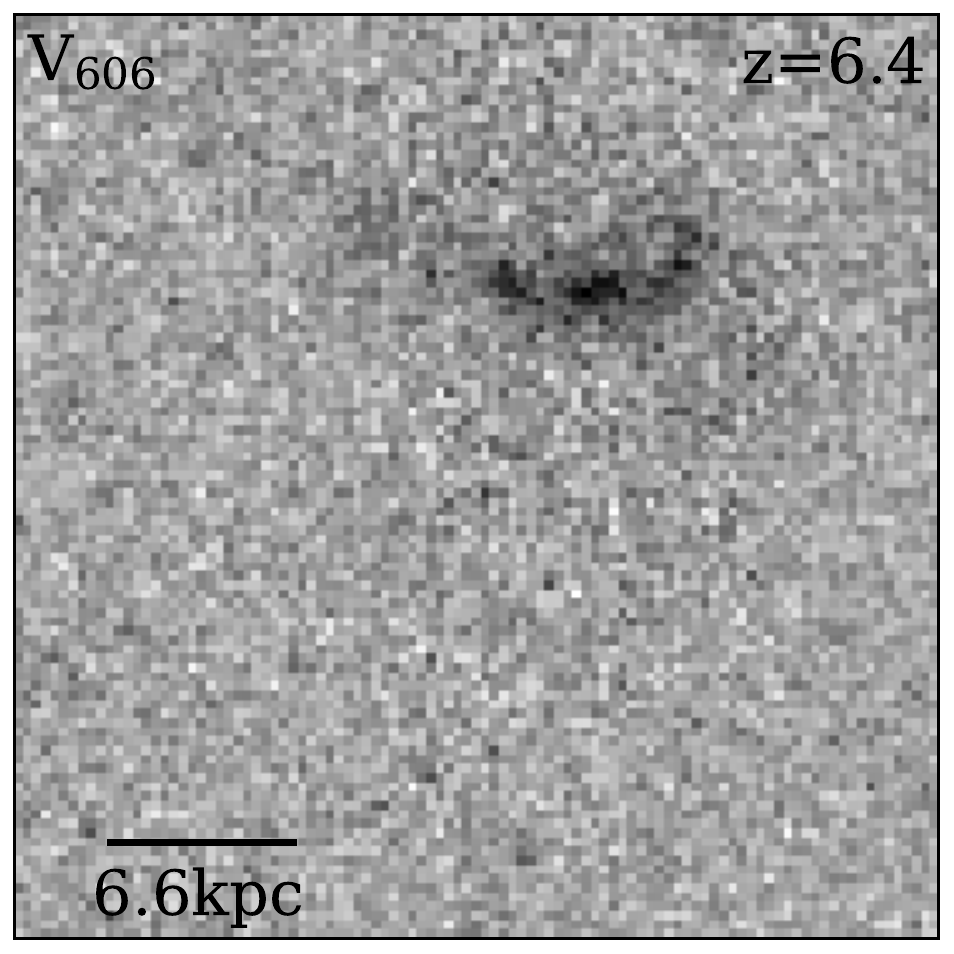}} &
\subfloat{\includegraphics[width = .23\textwidth]{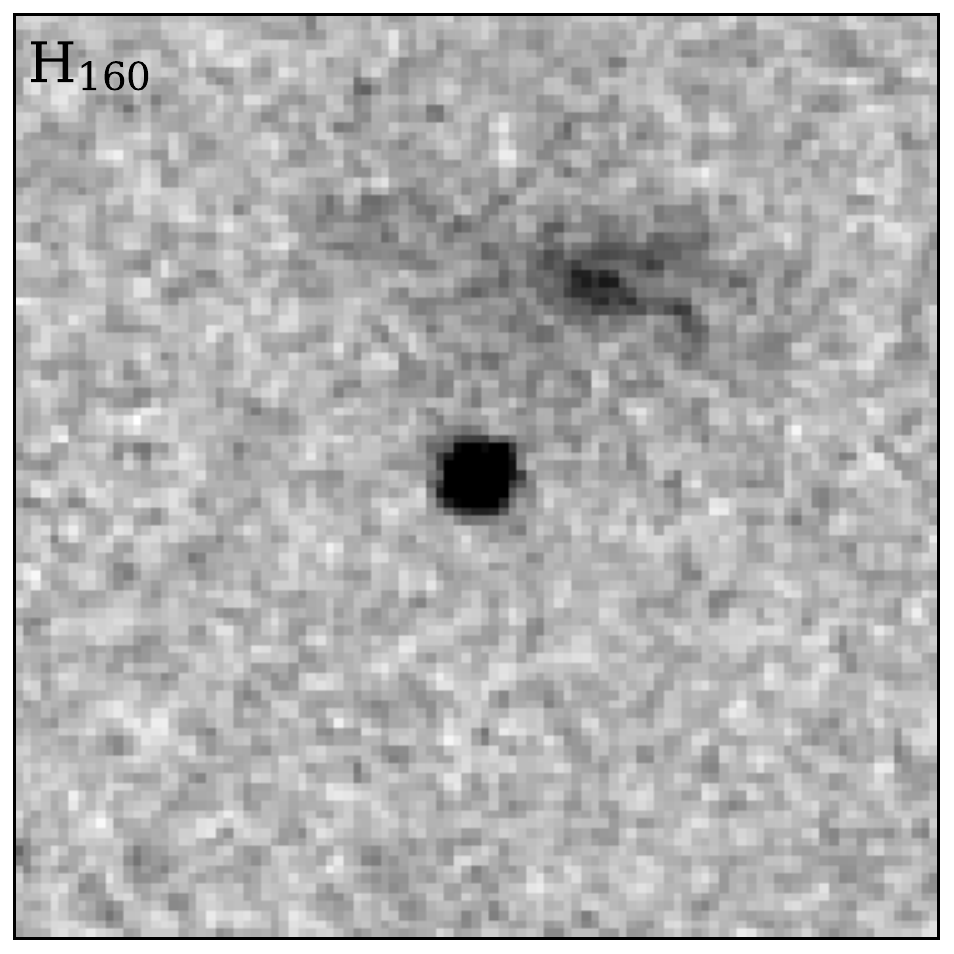}} &
\subfloat{\includegraphics[width = .23\textwidth]{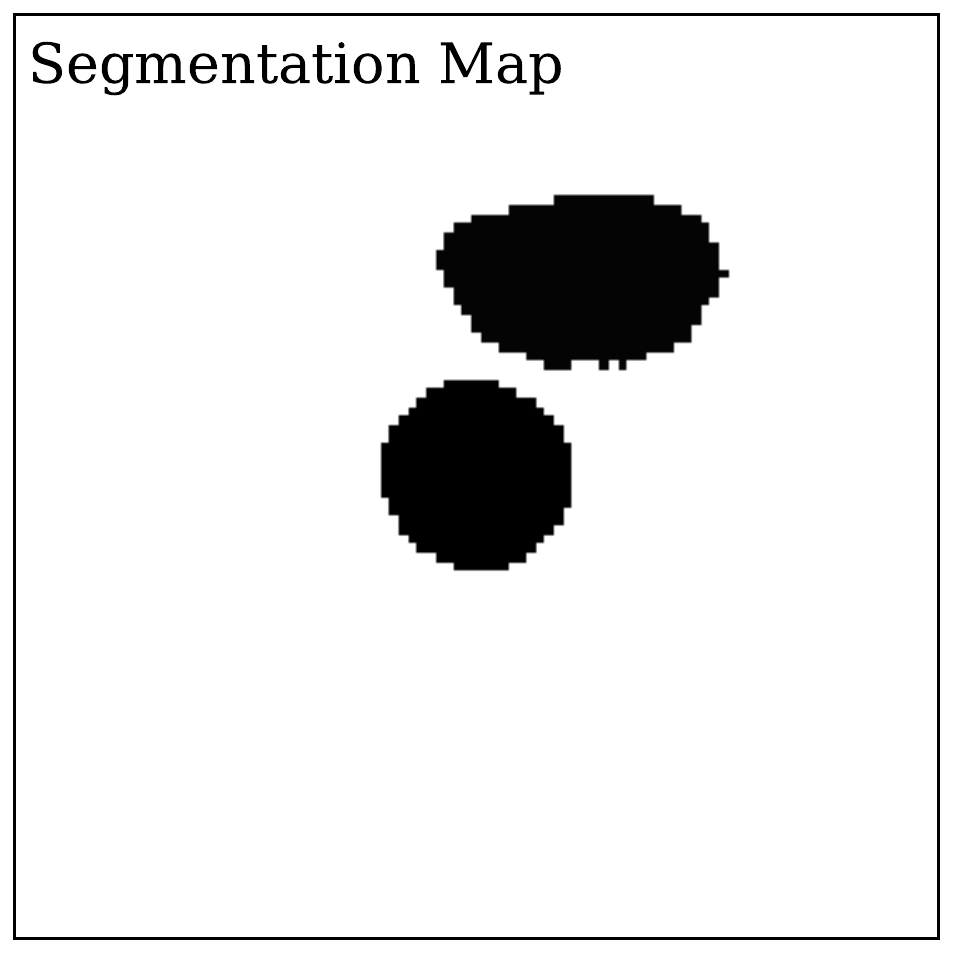}} &
\subfloat{\includegraphics[width = .23\textwidth]{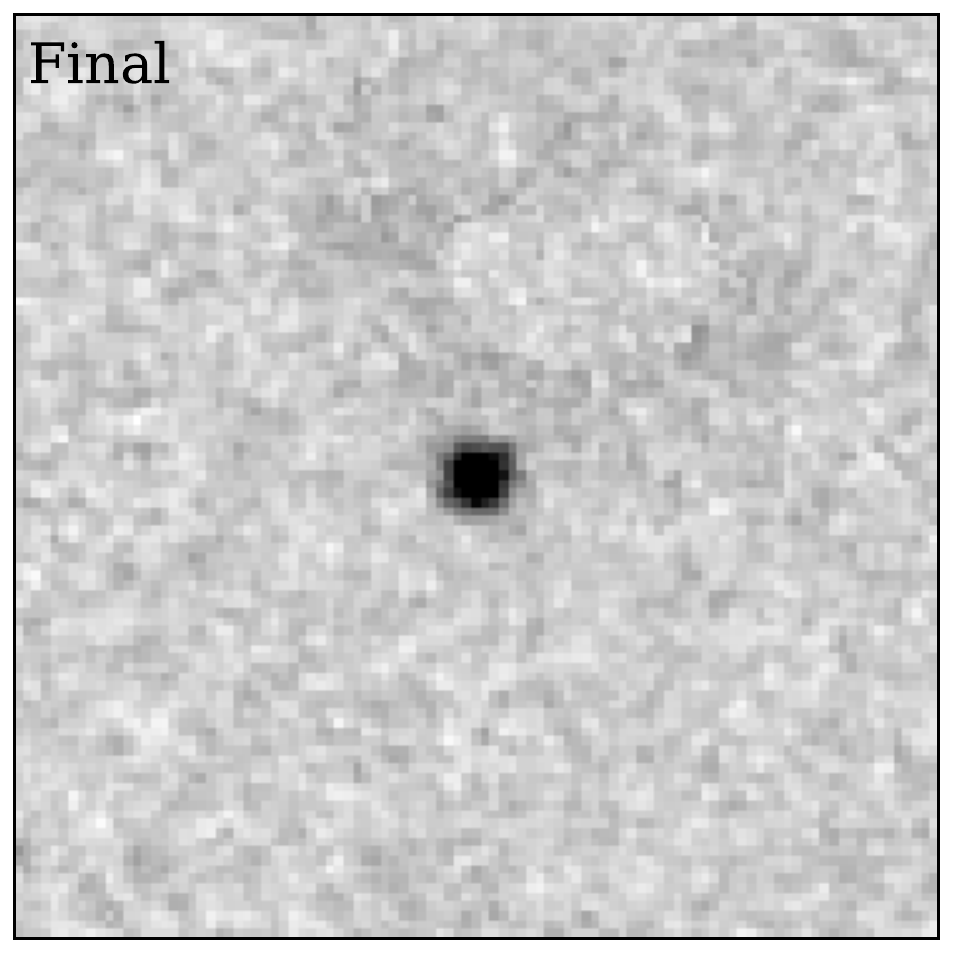}} \\
\subfloat{\includegraphics[width = .23\textwidth]{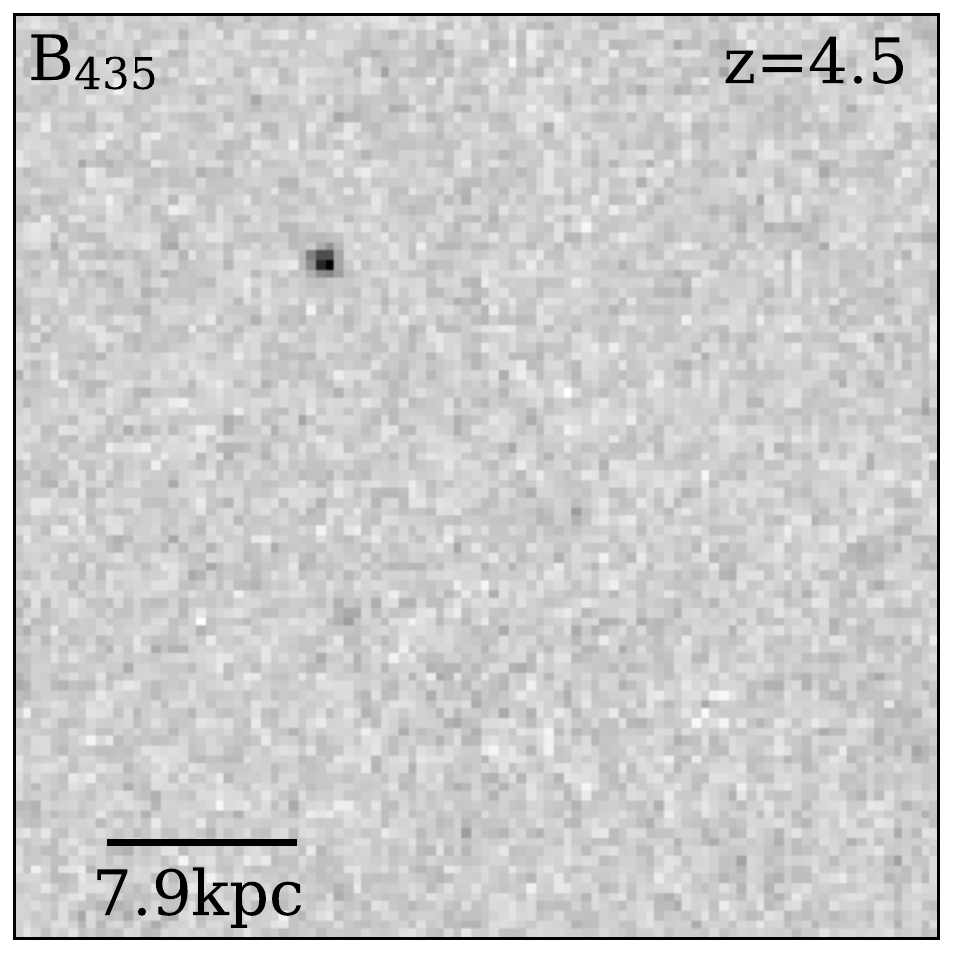}} &
\subfloat{\includegraphics[width = .23\textwidth]{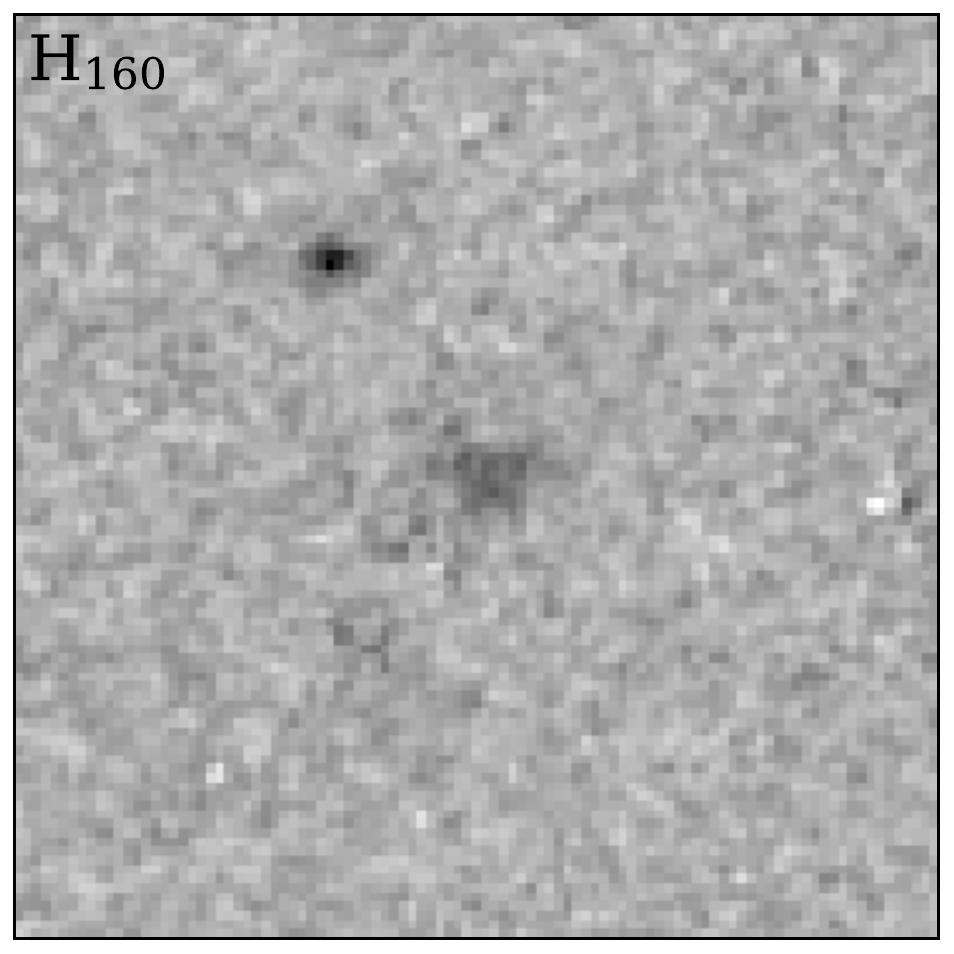}} &
\subfloat{\includegraphics[width = .23\textwidth]{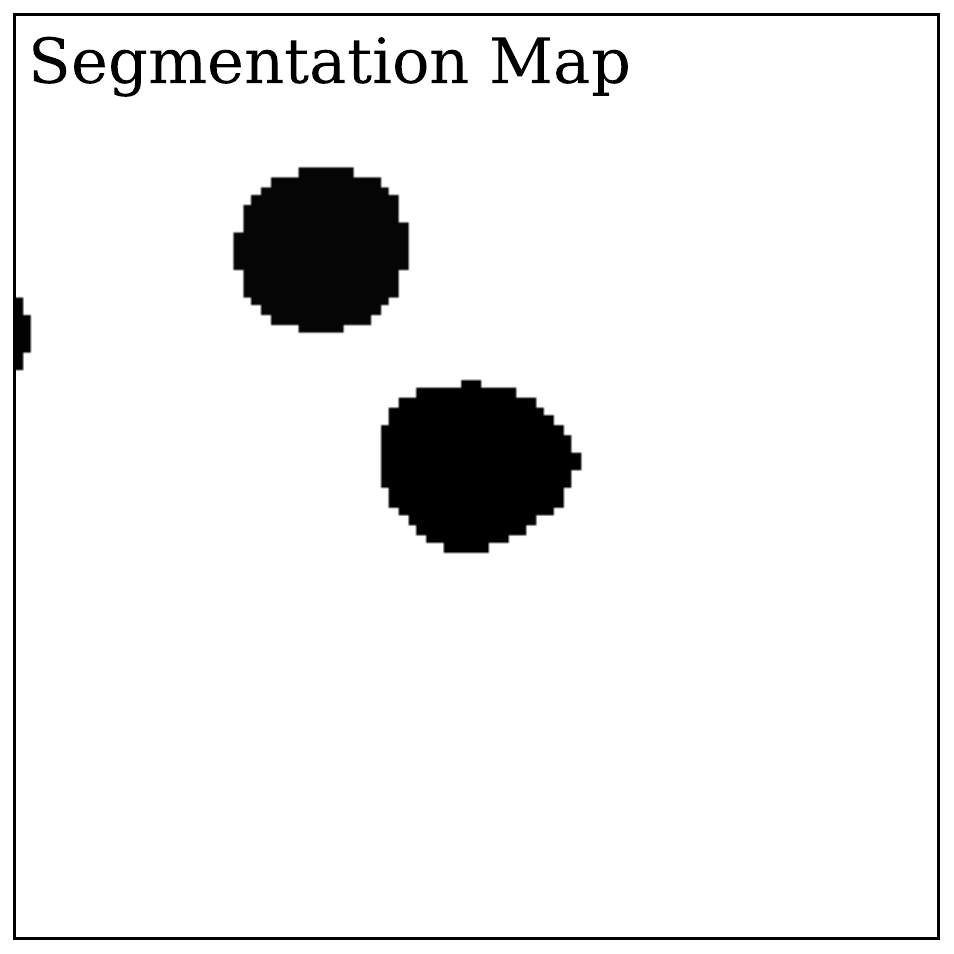}} &
\subfloat{\includegraphics[width = .23\textwidth]{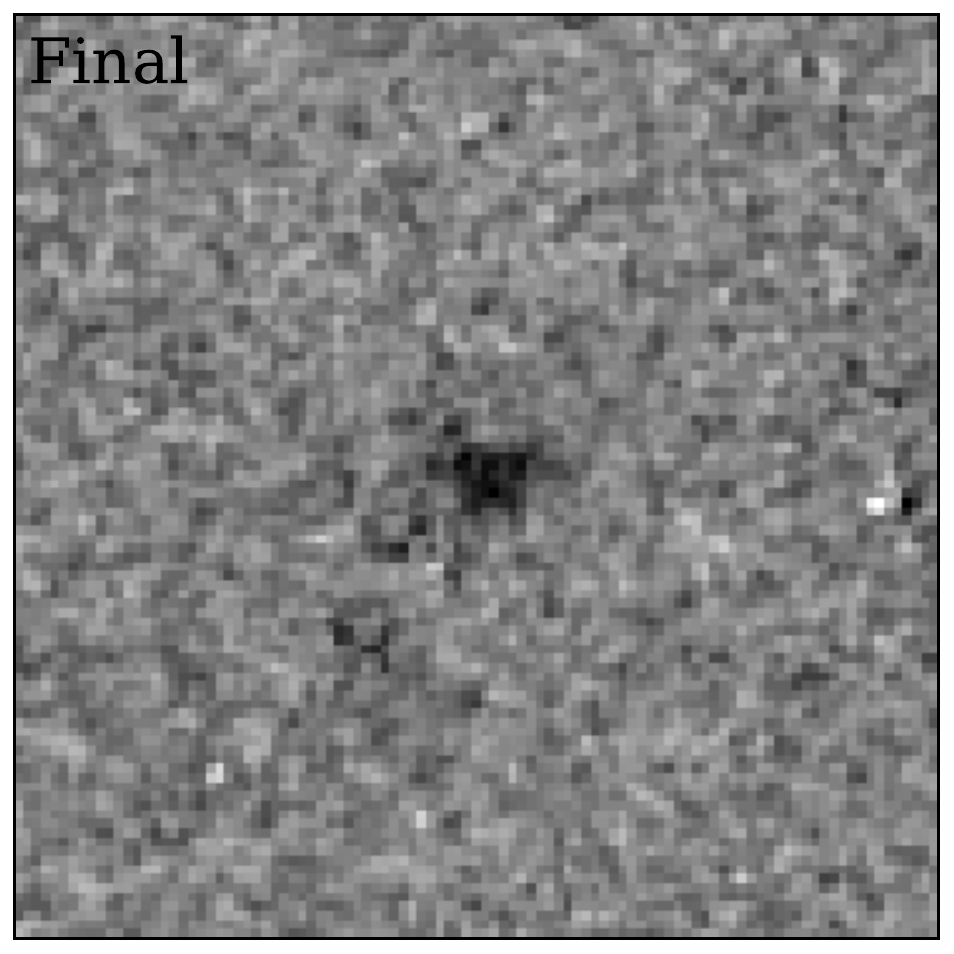}} \\
\subfloat{\includegraphics[width = .23\textwidth]{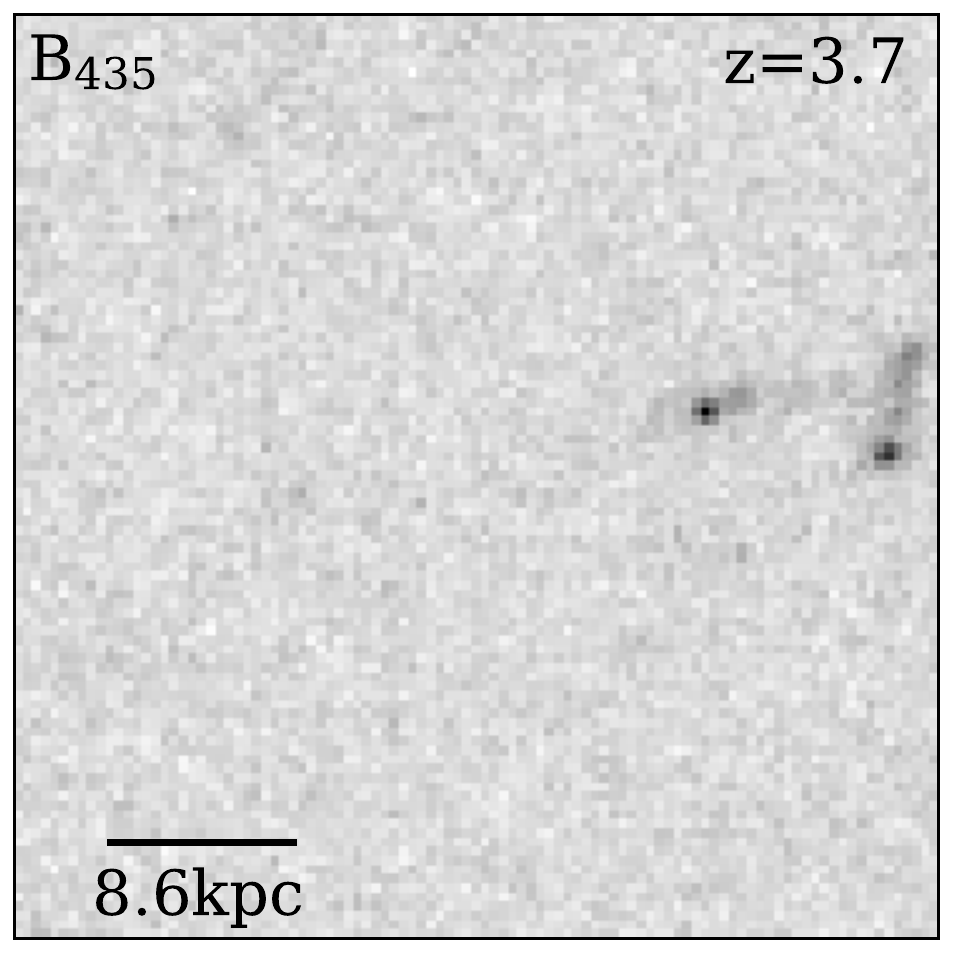}} &
\subfloat{\includegraphics[width = .23\textwidth]{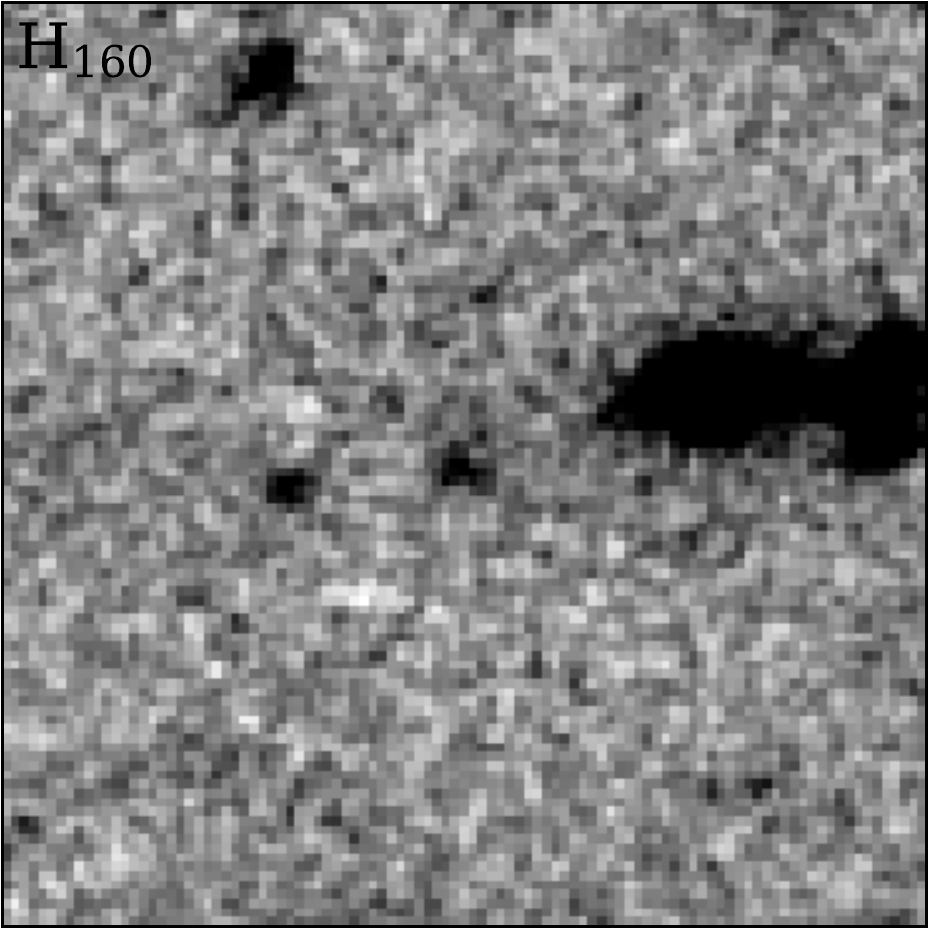}} &
\subfloat{\includegraphics[width = .23\textwidth]{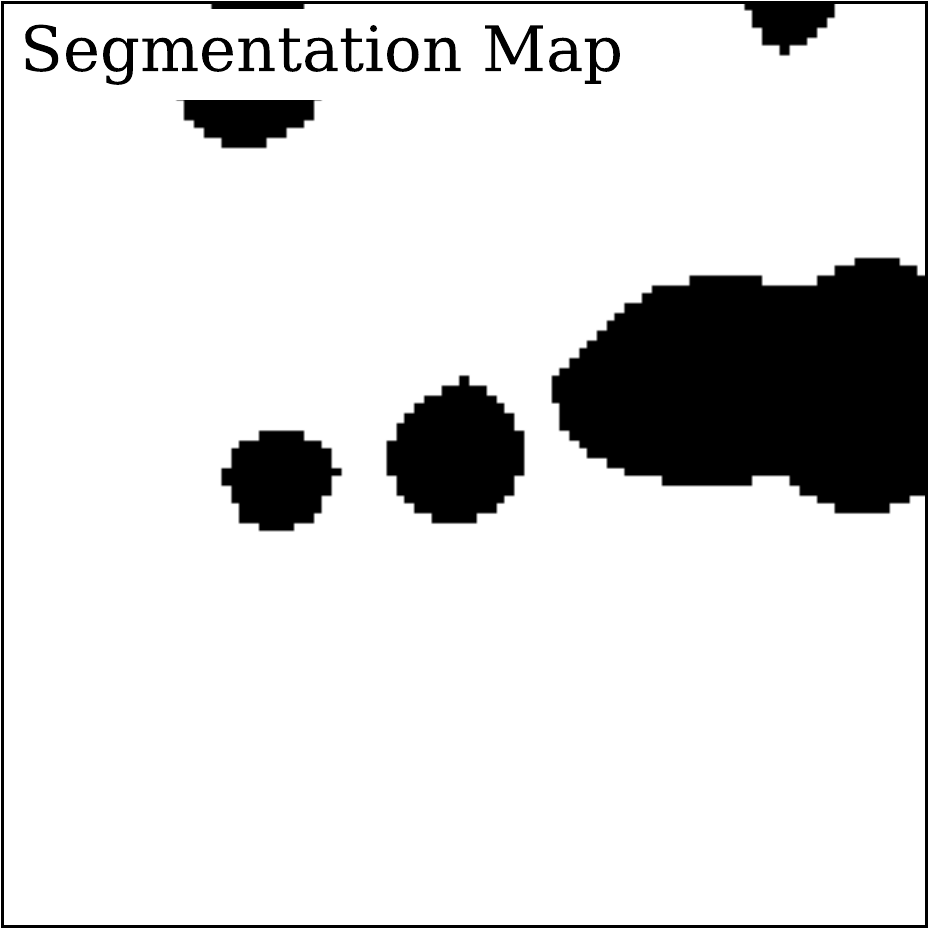}} &
\subfloat{\includegraphics[width = .23\textwidth]{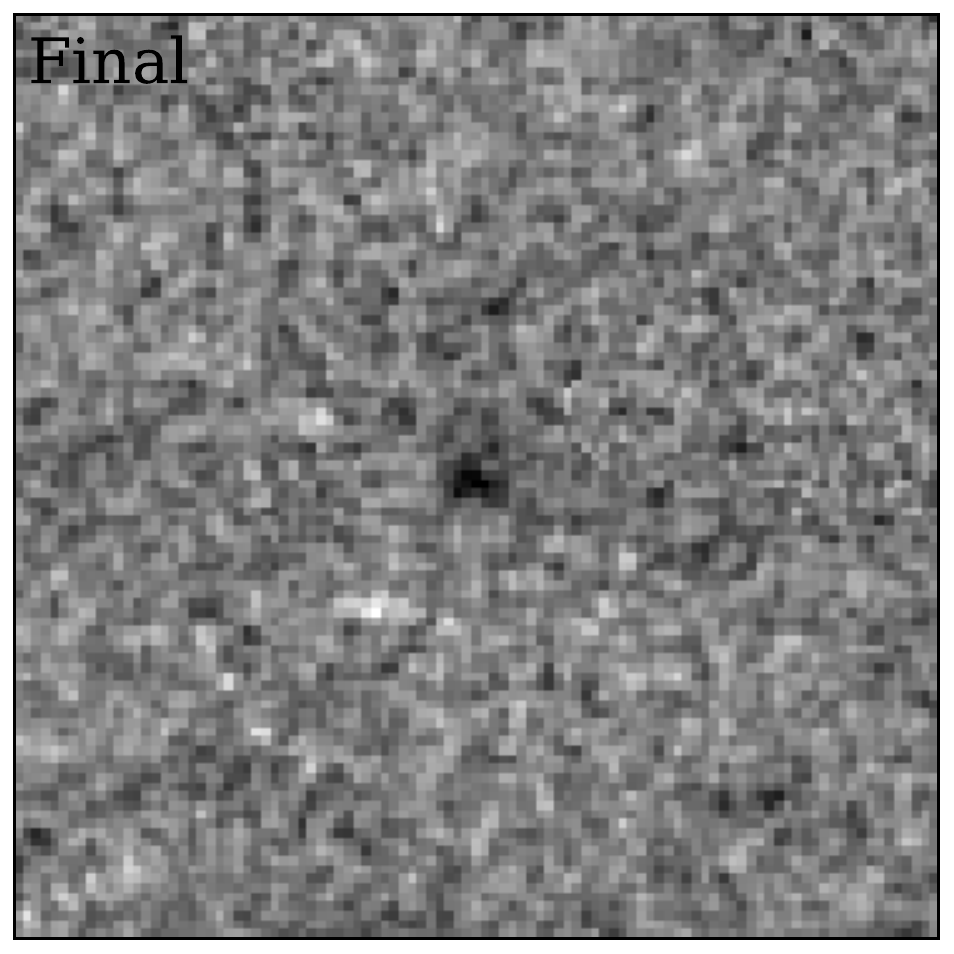}} \\
\end{tabular}
\caption{Examples of our image processing technique for four galaxies at redshifts of 7.0, 6.4, 4.5, and 3.7. Each column ($D_{i,j}^{raw}$, $O_{i,j}^{raw}$, $S_{i,j}$, and $O_{i,j}^{analysis}$) corresponds to the parameters of equation \ref{eq:imgproc}. The first column (left) shows the original V$_{606}$ or B$_{435}$ image showing the light below the Lyman-break rest-frame wavelength for the central galaxy's redshift, the second column shows the original H$_{160}$ band image, the third column shows the segmentation map corresponding to the optical rest-frame, while the fourth column (right) shows the result of the image processing whereby all galaxies that appear below the Lyman-break are removed (see equation \ref{eq:imgproc} for details). The field of view is 6'' on a side.}
\label{fig:img_process}
\end{figure*}

The basic technique was first used by \cite{steidel96} to find distant galaxies as unresolved objects in ground based imaging but can also be used in a 2-dimensional way to remove foreground and background galaxies for systems where the Lyman-break is visible within resolved imaging as with the Hubble Space Telescope. We call this '2-D Lyman-Break Imaging', an earlier version of which is described in \cite{conselice09}.

Initially, postage stamps measuring 6''$\times$6'' (100 pixels $\times$ 100 pixels) in size are created from a mosaic image of the field. The postage stamps contain the target object at the centre and contain other galaxies projected near the galaxy at different redshifts. In order to minimise contamination from the field objects, the target objects are isolated by removing these potentially contaminating objects. The following steps of our procedure are probably best demonstrated with images as shown in Figure \ref{fig:img_process}. To remove the foreground objects, the band corresponding to the Lyman-break and below is subtracted from the optical rest-frame image. The resulting image is then normalised by the optical rest-frame image. Maps of the pixels associated with the galaxy of interest are created such that the pixels corresponding to the central object are given a value of one and the pixels corresponding to the sky are given a value of 0. These are created by selecting pixels that have a value that is equal to or greater than three times the standard deviation of the background statistics. 

This map is used in combination with the segmentation map of the optical rest-frame image to remove areas of the sky that are identified as field objects. These removed areas and objects are then replaced with noise that has the same mean and standard deviation as the sky. Figure \ref{fig:img_process} shows this process for four galaxies at redshifts of 7.0, 6.4, 4.5 and 3.7. On the left we show the original V$_{606}$ or B$_{435}$ band images. Only the foreground objects are visible in these bands. In the second column, we show the optical rest-frame image for each of our sample galaxies where both the central object and potentially contaminating foreground objects are visible. In the third column, we show the corresponding segmentation map which highlights those pixels that are associated with the target galaxy and other objects. There are small differences between the objects appearing in the blue and red images however we always use the segmentation map that corresponds to the optical rest-frame such that these foreground objects are removed completely. 

On the right of Figure \ref{fig:img_process}, we show the result of the image processing where we have removed the foreground objects from the image. It is this final image in which we carry out our size analysis.

Our image processing technique can be described by the equation

\begin{equation}
    O_{i,j}^{analysis} = \left(\frac{O_{i,j}^{raw}-D_{i,j}^{raw}}{O_{i,j}^{raw}} \cdot S_{i,j}\right) + f(O_{i,j}^{raw, sky}) 
    \label{eq:imgproc}
\end{equation}

\noindent where $O_{i,j}^{raw}$ is the original optical rest-frame image or its substitute, $D_{i,j}^{raw}$ is the original drop-out image, $S_{i,j}$ is the segmentation map (shown in column 3 of Figure \ref{fig:img_process}), and $f(O_{i,j}^{raw, sky})$ is some function of the raw optical rest-frame image. The function $f(O_{i,j}^{raw, sky})$ creates an image in which the pixels corresponding to the central object are 0, the pixels corresponding to the sky are those of the raw optical rest-frame image, and the pixels corresponding to the field objects are noise that has the same mean and standard deviation of the sky. Both $O_{i,j}^{raw}$ and $D_{i,j}^{raw}$ must be in the same units.

We measure the sizes of our galaxies using the images produced using our 2-D Lyman-break method of removing foreground objects from the optical rest-frame images of our galaxy sample. This allows us to probe the rest-frame at $\sim$ 4000{\AA} wherever possible. The bands used for the image processing are shown in Table \ref{tab:imgprocessing}, along with the rest frame wavelength we probe at each redshift.

\begin{table}
\centering
\caption{The bands used to complete the image processing for each redshift in column 1. Column 2 gives the band corresponding to the optical rest-frame ($O_{i,j}^{raw}$), and Column 3 gives the band corresponding to the Lyman-break where applicable ($D_{i,j}^{raw}$). Column 4 gives the rest frame wavelength probed.}
  \begin{tabular}{cccc}
  \hline 
  \hline
  $z$ & $O_{i,j}^{raw}$ & $D_{i,j}^{raw}$ & $\lambda_{rest}$\\
  \hline
  1 & I$_{814}$ & - & 4070{\AA} \\
  2 & J$_{125}$ & - & 4170{\AA} \\
  3 & H$_{160}$ & - & 4000{\AA} \\
  4 & H$_{160}$ & B$_{435}$ & 3200{\AA} \\
  5 & H$_{160}$ & B$_{435}$ & 2670{\AA} \\
  6 & H$_{160}$ & V$_{606}$ & 2290{\AA} \\
  7 & H$_{160}$ & V$_{606}$ & 2000{\AA} \\
  \label{tab:imgprocessing}
  \end{tabular}

\end{table}

\subsection{Galaxy Sizes}

This work uses the Petrosian Radius ($R_{\textup{Petr}}(\eta)$) which is defined as the radius at which the surface brightness at a given radius is a particular fraction of the surface brightness within that radius \citep[e.g.][]{bershady00, conselice03b}. The concept of defining a size of a galaxy by the rate of change of light as a function of radius was first proposed by \cite{petrosian76} for cosmological uses. The radius measured depends on a defined ratio ($\eta(r)$) of surface brightness. $\eta(r)$ is defined as 

\begin{equation}
\eta(r) = \frac{I(r)}{\left\langle I(r) \right\rangle}
\end{equation}

\noindent where $I(r)$ is the surface brightness at radius $r$ and $\left\langle I(r) \right\rangle$ is the mean surface brightness within that radius. By this definition, $\eta(r)$ is 1 at the centre and 0 at large $r$ \citep{kron95}. The Petrosian radius at $\eta = 0.2$ contains at least 99$\%$ of the light within a given galaxy \citep{bershady00}. 

The Petrosian radius we use is determined using the CAS (concentration, asymmetry, and clumpiness) code \citep{conselice03b} which provides two measurements of size (Petrosian radius and half-light radius) along with the CAS parameters. The Petrosian radius differs from the half-light radius in that the former is a redshift independent measure of galaxy size. As such, the Petrosian radius of a particular galaxy would be, in principle, measured as the same no matter what redshift it was placed, whereas the half-light radius would potentially decrease as the redshift increases and outer light is lost. 

\begin{figure}
\includegraphics[width=0.475\textwidth]{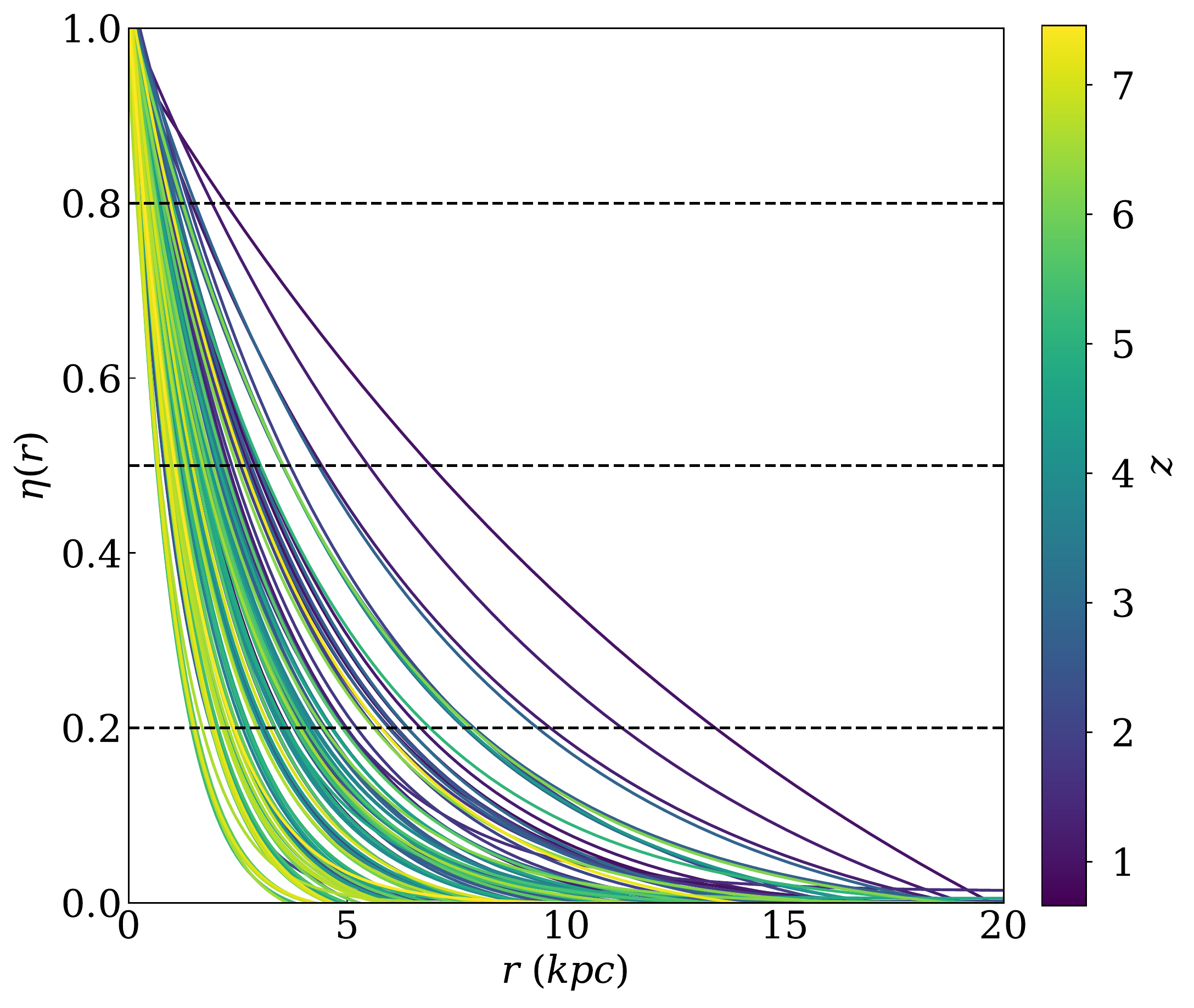}
\caption{Exponential fits of the form $\eta(r) = ae^{-cr} + d$ of the $\eta$ profiles for 98 random galaxies from our sample showing that on average, the higher redshift galaxies are smaller than those at lower redshift. The line colour corresponds to the redshift of each of the galaxies, as indicated by the colourbar on the right. The three horizontal lines indicate the locations of $\eta$ = 0.2, 0.5, and 0.8.}
\label{fig:etaprof}
\end{figure}

We however examine this assumption and correct for the effects of the PSF in the measurement of the Petrosian radii through simulating images and measuring radii in the same way as we do for our sample galaxies.  In Figure \ref{fig:etaprof}, we show how $\eta$ varies with radius $r$ for 98 random galaxies within our sample across a range of redshifts. We show that on average, those galaxies at the higher redshifts (yellow lines) are smaller in size than those at a given lower redshift (purple lines). The lines plotted are exponential fits of the $\eta$ profiles of the form 

\begin{equation}
  \eta(r) = ae^{-cr} + d.   
\end{equation}

\noindent The horizontal lines indicate the positions of the three $\eta$ values used throughout. 

\subsection{Simulations} \label{sec:sims}

\begin{figure}
\centering
\begin{tabular}{cccc}
\subfloat{\includegraphics[width = .23\textwidth]{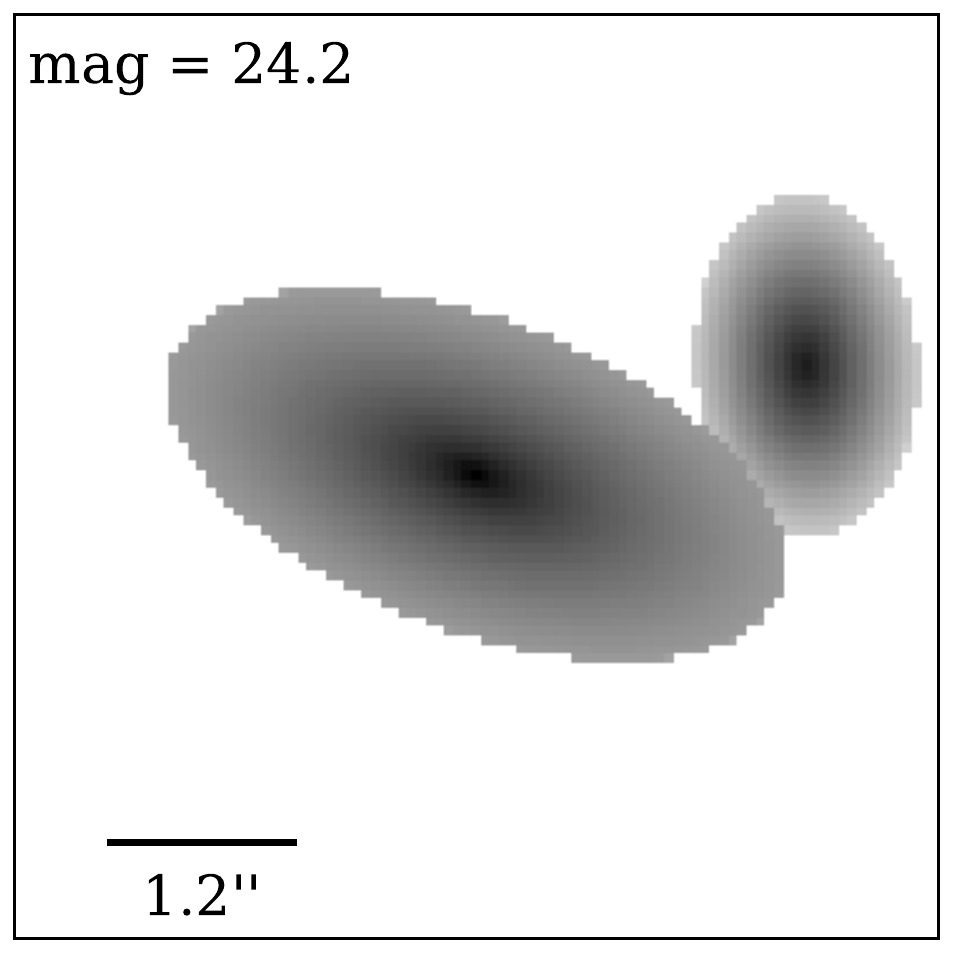}} &
\subfloat{\includegraphics[width = .23\textwidth]{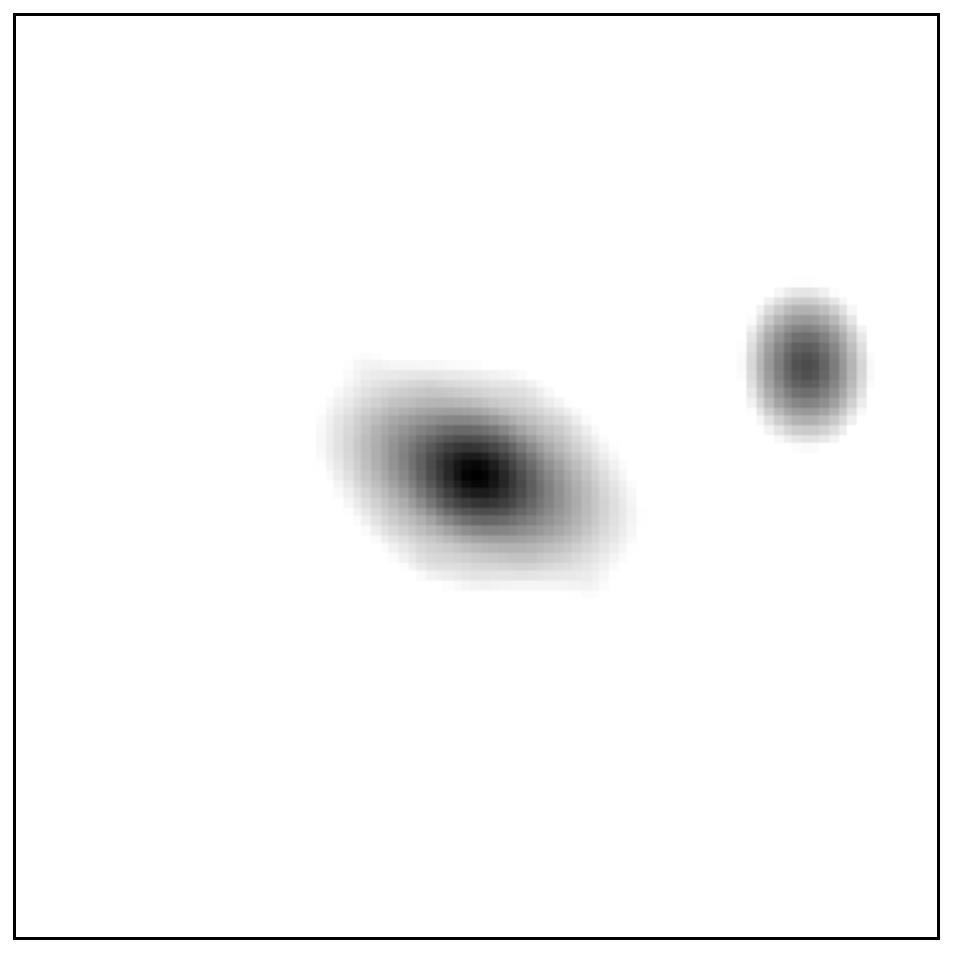}} \\
\subfloat{\includegraphics[width = .23\textwidth]{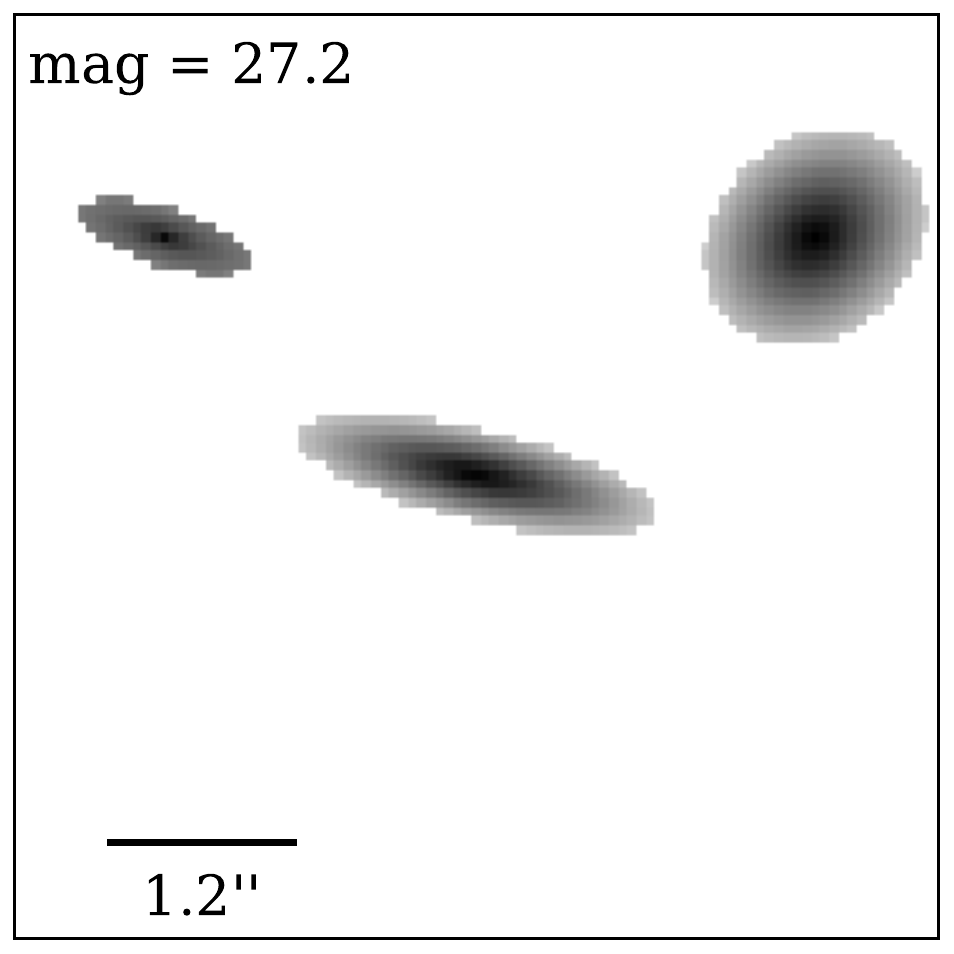}} &
\subfloat{\includegraphics[width = .23\textwidth]{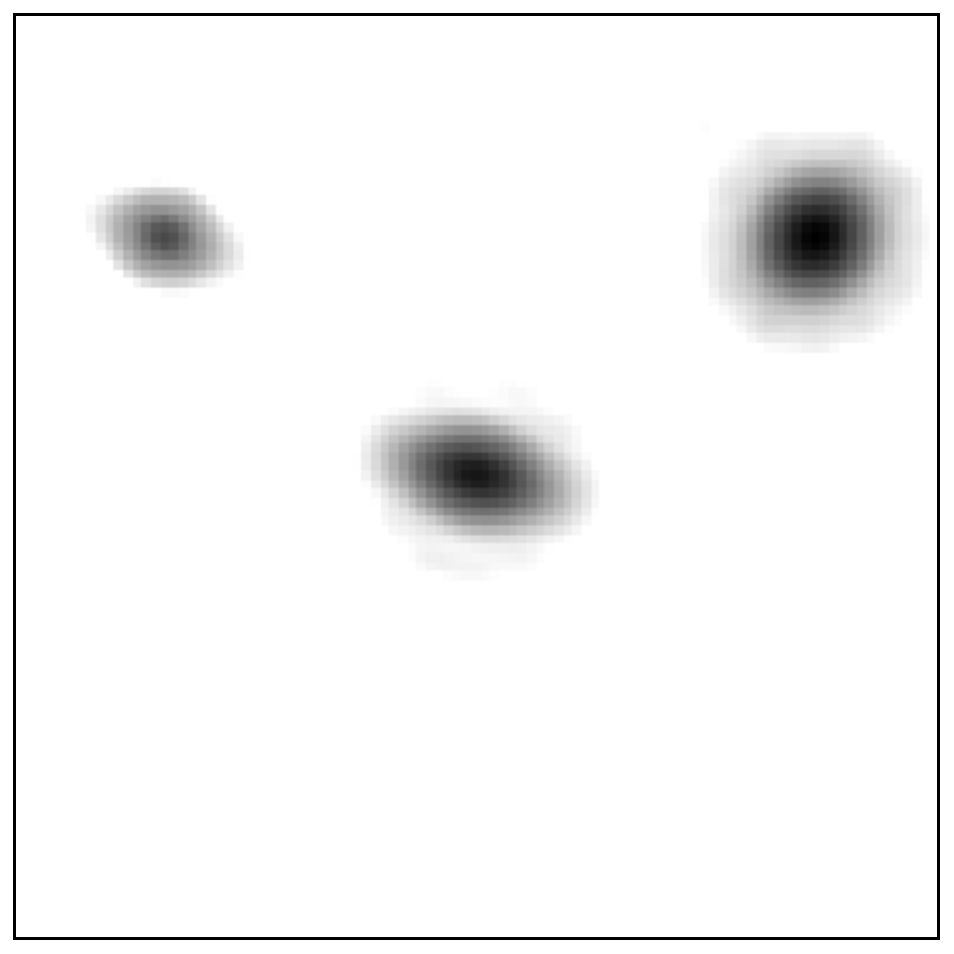}} \\
\subfloat{\includegraphics[width = .23\textwidth]{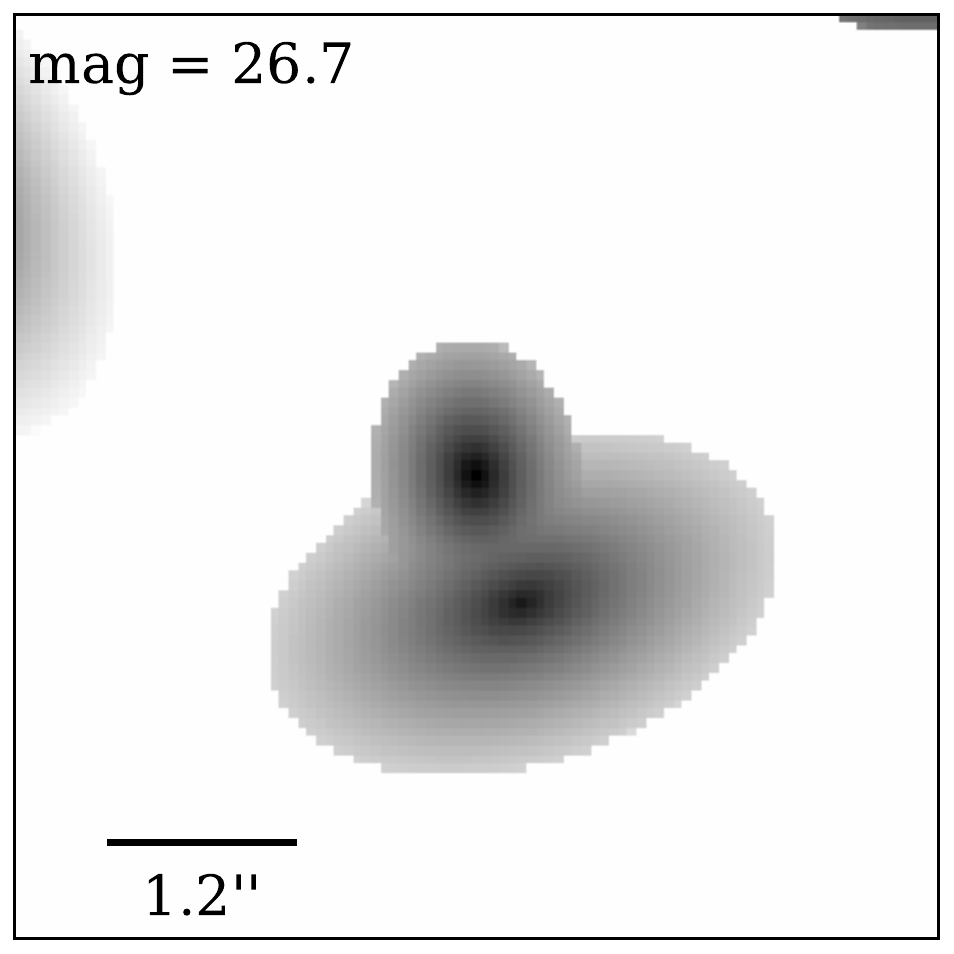}} &
\subfloat{\includegraphics[width = .23\textwidth]{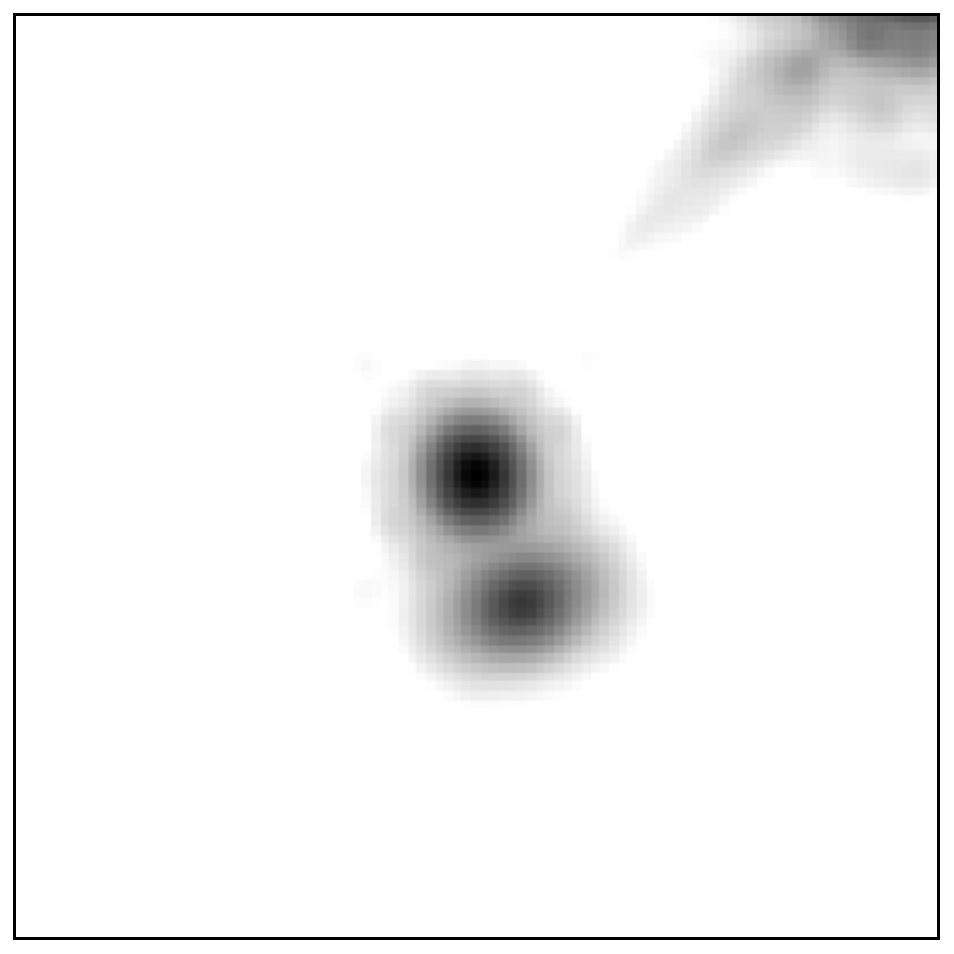}} \\

\end{tabular}
\caption{Examples of the simulated galaxies before the WFC3 PSF is applied (left) and after (right). These are idealized images which we use to correct the effects of PSF on our size measurements.}
\label{fig:sims}
\end{figure}

To determine how well we can measure galaxy sizes through Petrosian radii we follow the same method as \cite{bhatawdekar19} and simulate a sample of galaxies using the \textsc{IRAF} task, \textsc{mkobjects} in order to determine how much of a correction to the measured radii is required. The sample of 1912 simulated galaxies is uniformly distributed across the simulated field and a luminosity distribution of the form of a power law is applied to create a range of magnitudes. The simulated galaxies lie within a magnitude range of 21 to 30 and a size range of 2 to 42 pixels. We apply a range of surface brightness profiles to the sample of simulated galaxies with Sersic indices in the range $0.5 < n < 4$. The simulated galaxies are convolved using the WFC3 point spread function (PSF). We use the same PSF for each of the simulated galaxies due to the fact that any potential PSF variations do not make a significant impact at this level as we use it solely on the simulated images and not in any fits produced. After this image is created, \textsc{SExtractor} \citep{bertin96} is run on on the new image to detect the sources. A postage stamp measuring 100 pixels $\times$ 100 pixels of each object (pre- and post-convolution) is created and examples of the simulated galaxies can be seen in Figure \ref{fig:sims}. We show here the images before the WFC3 PSF has been applied on the left and the images after the PSF has been applied on the right. The CAS code we use for the real sample is then applied to this simulated sample to calculate the Petrosian radius of each of the objects in the same way we did for our original sample. We then compare the radii measured before applying the PSF to the radii measured after the PSF is applied for each value of $\eta$ (0.2, 0.5, and 0.8) and the relationship between the two is obtained through a linear fit. 

\begin{figure*}
\centering
\begin{tabular}{ccc}
\subfloat{\includegraphics[width = .3\textwidth]{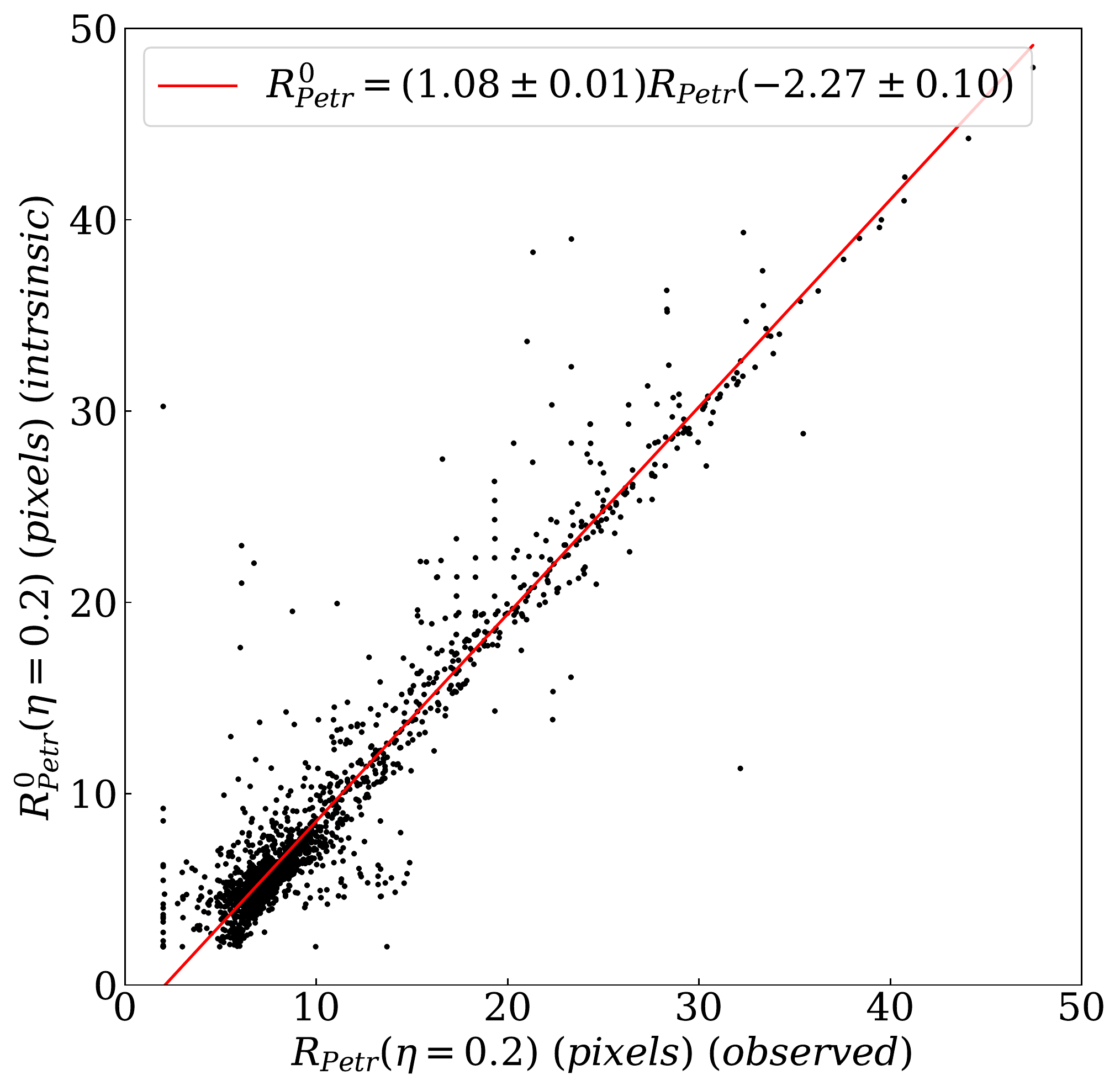}} &
\subfloat{\includegraphics[width = .3\textwidth]{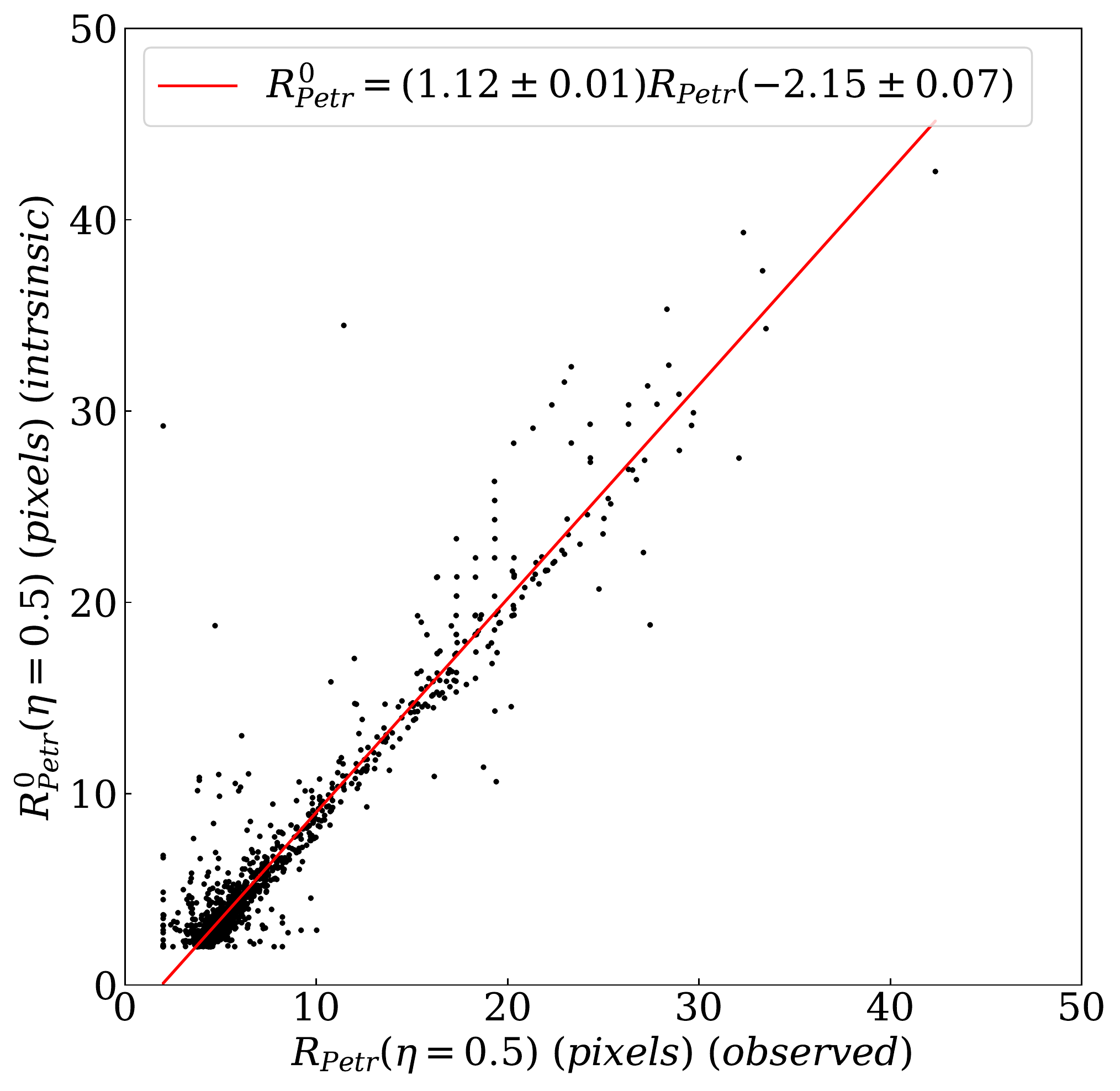}} &
\subfloat{\includegraphics[width = .3\textwidth]{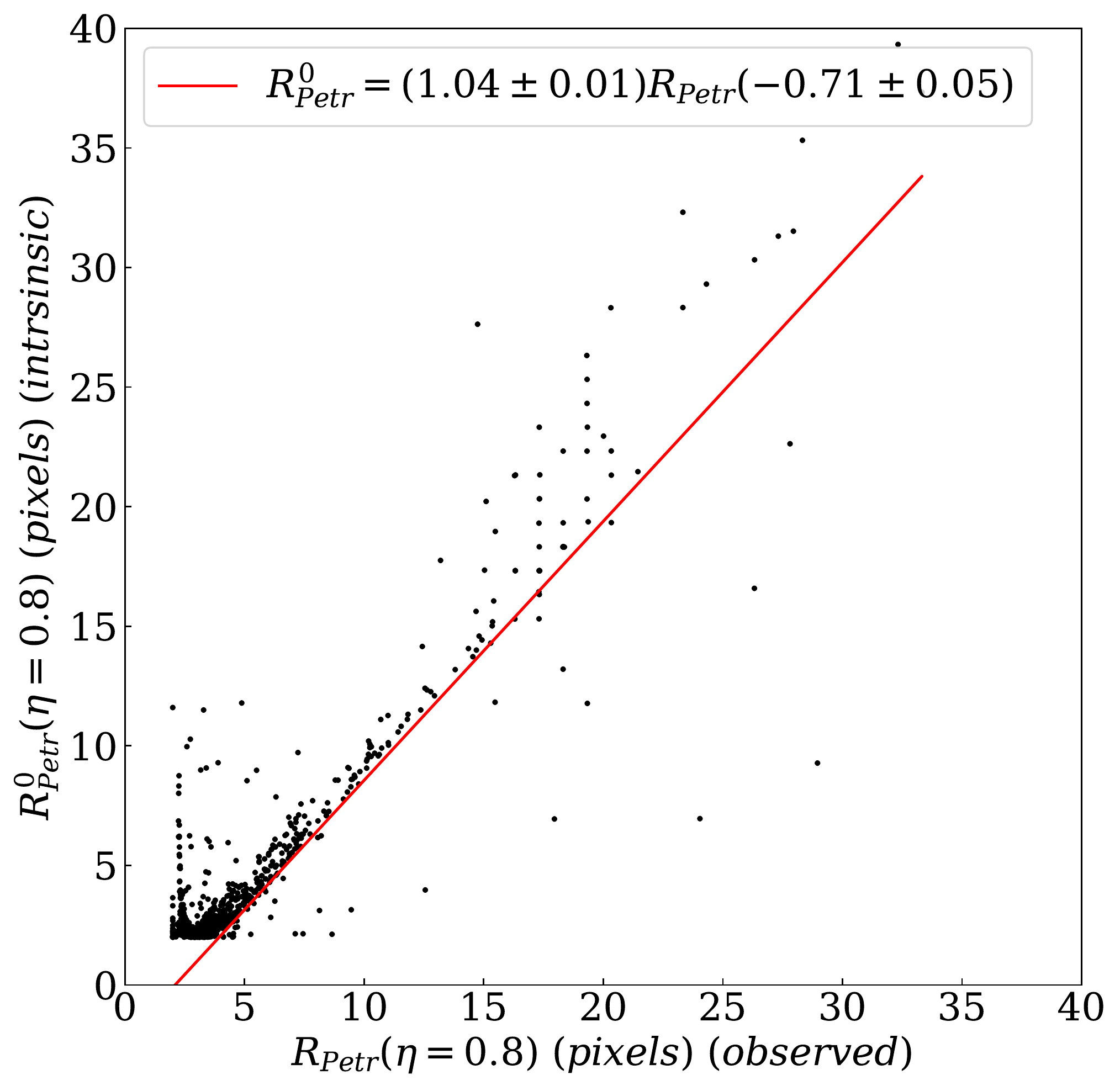}} \\
\end{tabular}
\caption{Relationship between the post-convolution sizes and the pre-convolution sizes measured on the simulated galaxies described in Section \ref{sec:sims} for $\eta = 0.2$ (left), $\eta = 0.5$ (middle), and $\eta = 0.8$ (right). Fits of the form $R_{intrinsic} = mR_{observed} + c$ for each Petrosian radius have been plotted as a red line. For $R_{\textup{Petr}}(\eta=0.2)$ we find $m = 1.08 \pm 0.01$ and $c = -2.27 \pm 0.10$. For $R_{\textup{Petr}}(\eta=0.5)$ we find $m = 1.12 \pm 0.01$ and $c = -2.14 \pm 0.10$. For $R_{\textup{Petr}}(\eta=0.8)$ we find $m = 1.04 \pm 0.01$ and $c = -0.71 \pm 0.05$. The post-convolution sizes are on average changed by 0.094'' for $\eta = 0.2$, 0.081'' for $\eta = 0.5$, and 0.036'' for $\eta = 0.8$.}
\label{fig:simsfit}
\end{figure*}

The relationship between the observed and intrinsic radii for each $\eta$ value for the simulated galaxies is shown in Figure \ref{fig:simsfit}. We show $R_{\textup{Petr}}(\eta=0.2)$ on the left, $R_{\textup{Petr}}(\eta=0.5)$ in the centre, and $R_{\textup{Petr}}(\eta=0.8)$ on the right. The linear fits are shown as a red line on each of the panels. This fit is then applied to our observed sample to correct the measured radii. On average, all three radii change by a factor of $\sim 0.8$ with the radii measured using $\eta = 0.8$ changing on average by 0.036'' (0.6 pixels), radii measured using $\eta = 0.5$ changing on average by 0.081'' (1.35 pixels), and radii measured using an $\eta = 0.2$ changing by 0.094'' (1.56 pixels) on average. The change in measured size for these simulated galaxies is very small. We find the best fit between the size before and after PSF convolution using the analytical form

\begin{equation}
    R_{intrinsic} = mR_{observed} + c.
\end{equation}

\noindent We henceforth correct our radii using these average values. 

\section{Results} \label{sec:results}

In this section we present the results achieved by measuring the sizes of our galaxies in two different sub-samples taken from the full sample as described in Section \ref{sec:data}; a mass selected sample where galaxies lie within a mass range of 10$^9$M$_{\odot}$$\leq$M$_*$$\leq$10$^{10.5}$M$_{\odot}$, and a number density selected sample where galaxies are selected using a constant number density of 1$\times$10$^{-4}$ Mpc$^{-3}$. Where reference is made to a galaxy's size, this is taken to be the measured Petrosian radius.

\subsection{Rest-Frame Wavelength and Biases}

The appearance of a galaxy depends greatly on its rest-frame wavelength, and a galaxy can have a different morphological and quantitative classification at different wavelengths \citep{windhorst02, taylormager07, mager18}. This is due to the fact that different wavelengths probe different aspects of a galaxy,  with bluer light probing star formation and redder light probing the older existing populations of stars. The young stars can often have a distribution which is quite different from the older stars, and this needs to be accounted for if we want to measure galaxy sizes at intrinsically different rest-frame wavelengths.

This is also true of the measured surface brightness, and therefore the measured Petrosian radii. This effect is more prominent at lower redshifts when the star formation has dropped significantly. At low redshift, there is little star formation therefore galaxies at this epoch appear less luminous in the UV and therefore are often smaller in the UV than those at higher redshifts where more star formation occurs \citep[e.g.][]{hopkinsbeacom06}. Furthermore, it has been shown that whilst the rest-frame UV and optical structures of galaxies are often significantly different in the local universe, this is not true at high redshifts where galaxies are in many ways extremely similar in terms of structure in the rest-frame UV and optical \citep[e.g.][]{papovich03, papovich05, conselice05, conselice11a}. It has also been shown that the measured size of a galaxy does not depend on the observed wavelength to the first order, even after correcting for surface brightness dimming and PSF broadening \citep{ribeiro16}. Therefore we are able to use images that correspond to the UV rest-frame at high redshifts, particularly in the case of the most massive galaxies as the variation is not so significant for this population \citep{cassata10}.

We thus measure the sizes of galaxies in the optical rest-frame at $\lambda \cong 4000{\AA}$ where possible. However, this is not possible for galaxies at $z > 3$ where we are forced to probe galaxy sizes in progressively bluer wavelengths down to the UV. To determine the bias resulting from this we compare the sizes measured in the observed rest-frame ($\lambda = 4000{\AA}$) to the intrinsic rest-frame  UV at $\lambda \sim  2200{\AA}$. We do this test at the lower redshifts at $1 < z < 5$ where we have both a rest-frame optical and rest-frame UV morphology. What we find when we do this is that the sizes at both wavelengths are approximately equal. To show this we fit a straight line of the form

\begin{equation}
    R_{\lambda_1} = mR_{\lambda_2} + c 
    \label{eq:restframe}
\end{equation}

\noindent to the results where $R_{\lambda_1}$ is the Petrosian radius at $\eta = 0.2$ measured in the intrinsic bluer rest-frame and $R_{\lambda_2}$ is the Petrosian radius at $\eta = 0.2$ measured in the redder rest-frame. The fits for each redshift can be found in Table \ref{tab:rffits}. We only include results up to $z = 5$ due to the availability of bands corresponding to the appropriate rest-frame. Table \ref{tab:rffits} also gives the bands used to compare the observed and intrinsic sizes. The mean difference between these two wavelengths ($\overline{\delta R}$) is shown in Column 6 of Table \ref{tab:rffits}. This difference has been normalised by the sum of the sizes measured at the two wavelengths. These values are extremely small, showing that the size measurements made at the bluer rest-frame are similar in magnitude to those made at the redder rest-frame wavelength.

\begin{table*}[!ht]
    \centering
        \caption{Slopes (column 2) and y-intercept (column 3) of the observed rest-frame size ($R_{\lambda_1}$) and intrinsic rest-frame size ($R_{\lambda_2}$) measurement fits (given by equation \ref{eq:restframe}), including errors for each redshift (column 1). Columns 4 and 5 give the bands corresponding to the observed and intrinsic rest-frame wavelengths respectively. Column 6 gives the mean difference between the size measured in the observed and intrinsic rest-frame wavelengths. This difference has been normalised by $R_{\lambda_1}$. $R_{\lambda_1}$ and $R_{\lambda_2}$ are the Petrosian radii measured at $\eta = 0.2$.}
    \begin{tabular}{cccccc}
    \hline 
    \hline
        $z$ & $m$ & $c$ & $\lambda_2$ & $\lambda_1$ & $\lvert\frac{\overline{\delta R}}{R_{\lambda_1}}\rvert$\\
        \hline
         1 & 0.5359 $\pm$ 0.0001 & 5.0641 $\pm$ 0.0053 & I$_{814}$ & B$_{435}$ & 0.06 \\
         2 & 0.6488 $\pm$ 0.0001 & 5.4437 $\pm$ 0.0044 & J$_{125}$ & V$_{606}$ & 0.63 \\
         3 & 0.6904 $\pm$ 0.0002 & 4.7803 $\pm$ 0.0156 & H$_{160}$ & z$_{850}$ & 0.41 \\
         4 & 0.4584 $\pm$ 0.0013 & 2.9556 $\pm$ 0.0635 & H$_{160}$ & Y$_{105}$ & 0.11 \\
         5 & 0.5941 $\pm$ 0.0015 & 2.3206 $\pm$ 0.0685 & H$_{160}$ & J$_{125}$ & 0.07 \\
    \end{tabular}

    \label{tab:rffits}
\end{table*}

\subsection{Redshift-Size Relation}

\begin{figure*}
\centering
\begin{tabular}{ccc}
\subfloat{\includegraphics[width = .3\textwidth]{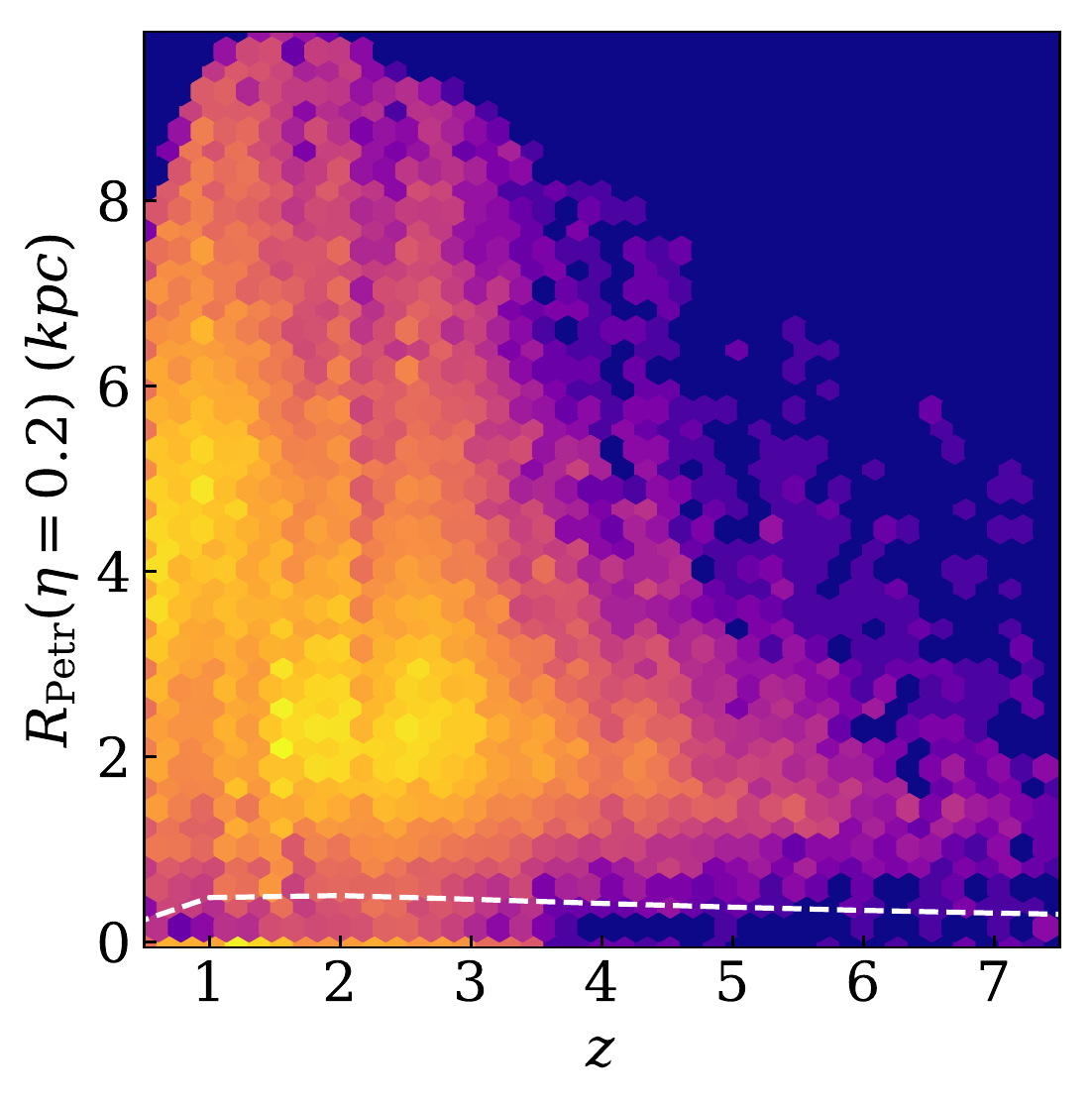}} &
\subfloat{\includegraphics[width = .3\textwidth]{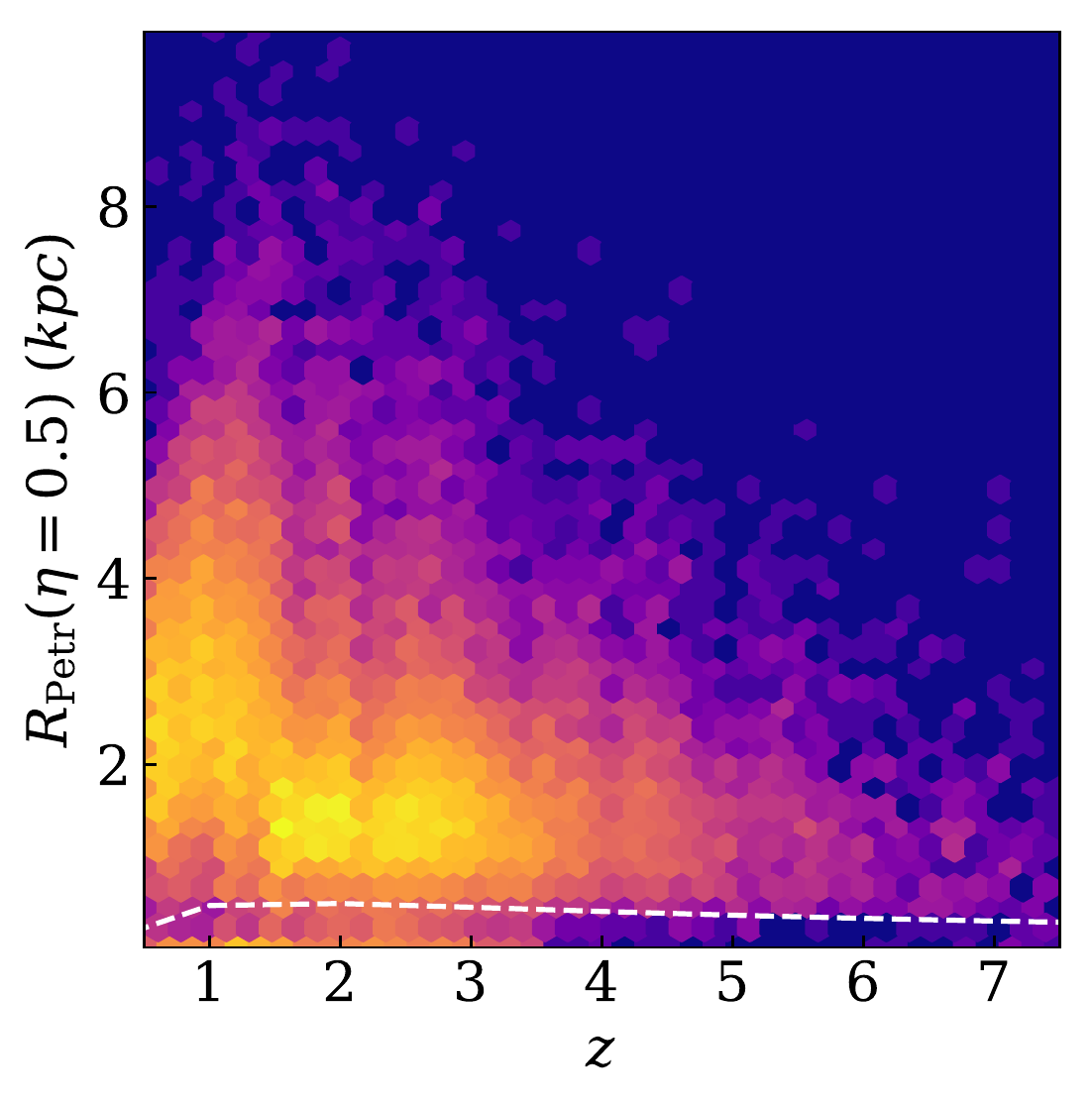}} &
\subfloat{\includegraphics[width = .31\textwidth]{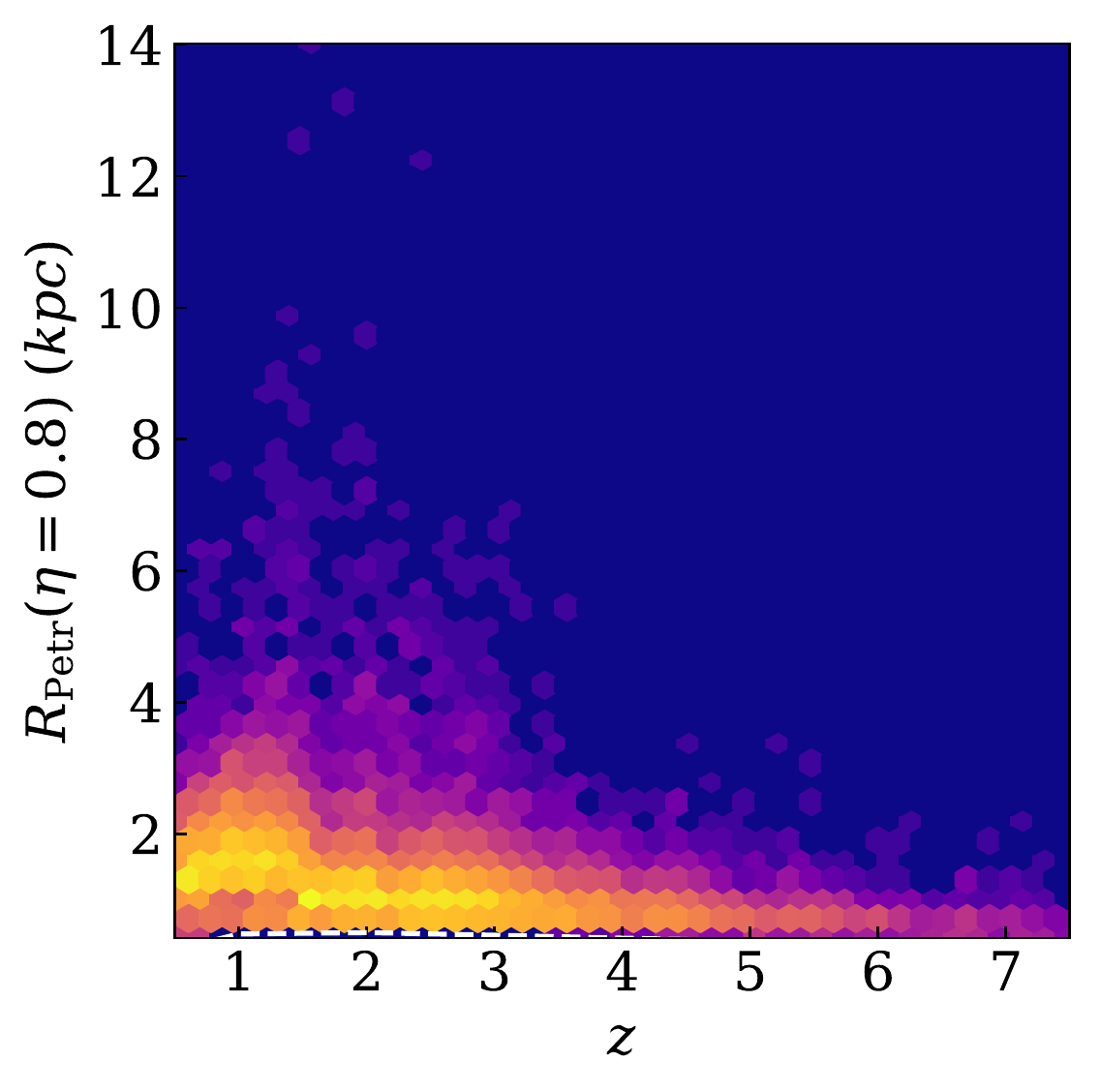}} \\
\end{tabular}
\caption{Galaxy size-redshift distribution for the full sample of galaxies for three different $\eta$ values. Left: $\eta = 0.2$. Middle: $\eta = 0.5$. Right: $\eta = 0.8$. The yellow regions show the highest density of points and purple the lowest. The pixel size is shown by the the white dashed line.}
\label{fig:rzdistribution}
\end{figure*}

By studying our full sample, we are able to see the effect redshift has on the size of galaxies for the full mass range. The distribution of the corrected Petrosian radii with redshift of the total sample size at three different $\eta$ values can be seen in Figure \ref{fig:rzdistribution} with $\eta = 0.2$ on the left, $\eta = 0.5$ in the middle, and $\eta = 0.8$ on the right. We represent the pixel size as a white dashed line. From the definition of the Petrosian radius, $\eta = 0.2$ corresponds to a measurement made near the outer edge of a galaxy, and $\eta = 0.8$ corresponds to a measurement made in the inner regions of a galaxy. The evolution of each with redshift changes in a similar way in that the there are more galaxies at larger radii at lower redshifts than seen at higher redshifts. However, the values of the radii differ such that the $\eta = 0.2$ values are typically much larger than those of $\eta = 0.8$ as expected.

\subsection{Mass-Size Relation}

\begin{figure*}[ht!]
\includegraphics[width=\textwidth]{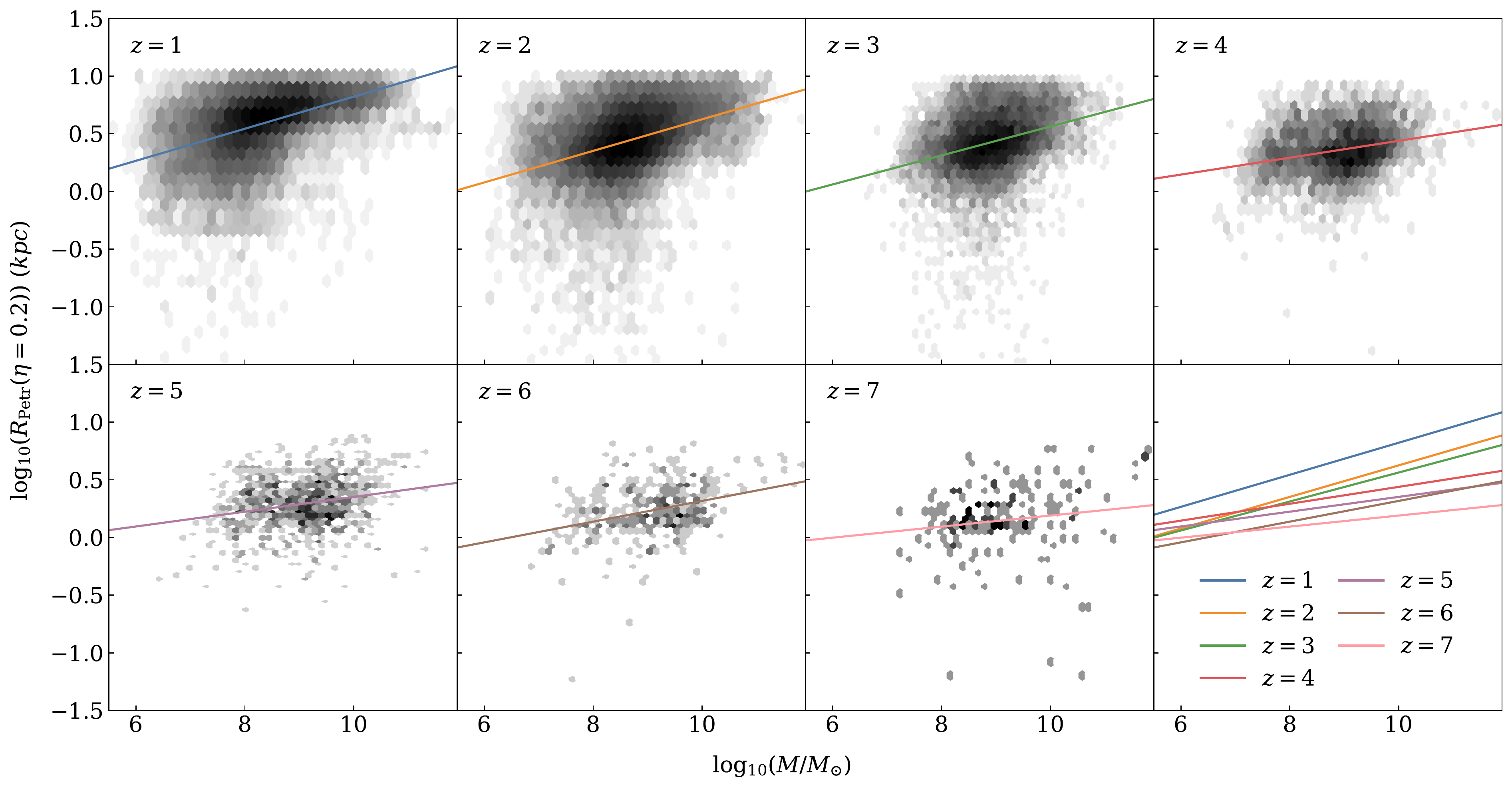}
\caption{Galaxy stellar mass-size distributions for the full sample of galaxies. The size measurement is given by $R_{\textup{Petr}}(\eta = 0.2)$. Each panel shows the distribution of a different redshift along with a line of best fit given in equation \ref{eq:massr}. In the final panel, we show the fitted mass-size relation for each redshift using the same coloured lines as in the individual panels. The gradient of each best fit is positive however it decreases as redshift increases. The best fit parameters are given in Table \ref{tab:massfits}.}
\label{fig:rmass}
\end{figure*}

From the full sample of galaxies we are able to determine the mass-size relation as a function of redshift. In Figure \ref{fig:rmass}, we show the mass-size distribution for each redshift bin. We use $R_{\textup{Petr}}(\eta = 0.2)$ as a measure of size in this case. Each panel corresponds to a different redshift, showing the density of the mass-size distribution and a line of best fit. In the final panel, we show the lines of best fit for all seven redshift bins, each in the same colour as their individual panels. The lines of best fit are of the form 

\begin{equation}
    \textup{log}_{10}(R_{\textup{Petr}}(\eta = 0.2)) = m\textup{log}_{10}(M/M_{\odot}) + c 
    \label{eq:massr}
\end{equation}

\noindent and the parameters for each redshift bin are shown in Table \ref{tab:massfits}. We see that the slope of the fit ($m$) decreases as redshift increases, showing that the sizes of the galaxies at lower redshifts have a greater dependence on their masses compared to those galaxies at higher redshifts. We also find that the intercept ($c$), on average, decreases as $z$ increases showing an evolution in size.

\begin{table}
\centering
\caption{The parameters determined for the mass-size relation (shown in Figure \ref{fig:rmass}) for each redshift bin. The fits are given in the form $\textup{log}_{10}(R_{\textup{Petr}}(\eta = 0.2)) = m\textup{log}_{10}(M/M_{\odot}) + c$. Where the error is given as 0, it is negligible in comparison to the value of the parameter.}
  \begin{tabular}{ccc}
  \hline 
  \hline
  $z$ & $m$ & $c$ \\
  \hline
  1 & 0.15$\pm$0.00 & -0.75$\pm$0.00 \\
  2 & 0.14$\pm$0.00 & -0.70$\pm$0.00 \\
  3 & 0.14$\pm$0.00 & -0.75$\pm$0.00 \\
  4 & 0.07$\pm$0.00 & -0.28$\pm$0.00 \\
  5 & 0.06$\pm$0.00 & -0.25$\pm$0.01 \\
  6 & 0.09$\pm$0.00 & -0.55$\pm$0.02 \\
  7 & 0.05$\pm$0.00 & -0.29$\pm$0.07 \\
  \label{tab:massfits}
  \end{tabular}

\end{table}

\subsection{Mass Selected Sample}

Here we present the results of the analysis of a mass selected sample (10$^9$M$_{\odot}$$\leq$M$_*$$\leq$10$^{10.5}$M$_{\odot}$) of 14,015 galaxies taken from the full sample. This mass range is chosen for completeness \citep{duncan14}. A mass selected sample allows us to remove biases present in the full sample due to the detection limits of the surveys. 

\begin{figure}[!ht]
\includegraphics[width=0.475\textwidth]{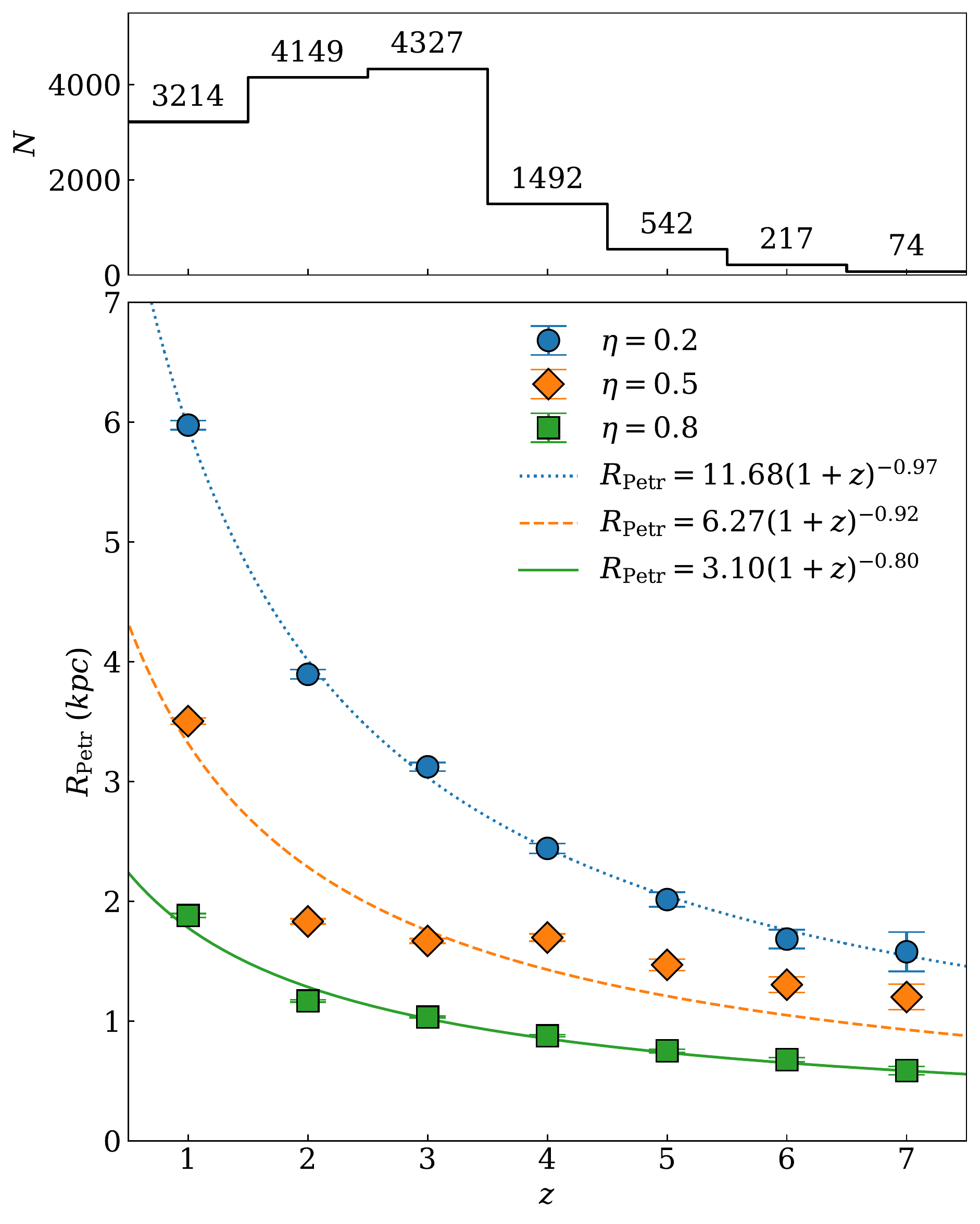}
\caption{Top: Histogram showing the distribution and number of galaxies in each redshift bin. Bottom: Evolution of the average Petrosian radius through redshift for the mass limited sample where a mass-cut of 10$^9$M$_{\odot}$$\leq$M$_*$$\leq$10$^{10.5}$M$_{\odot}$ is applied. Each point is the median value for each redshift bin with the error bars showing the standard error. Blue circles show how $R_{\textup{Petr}}(\eta=0.2)$ changes, orange diamonds show $R_{\textup{Petr}}(\eta=0.5)$, and green squares show $R_{\textup{Petr}}(\eta=0.8)$. By fitting a power-law relation to the median sizes, we find $R_{\textup{Petr}}(\eta=0.2)=11.68(1+z)^{-0.97\pm0.03}$ (blue dotted line), $R_{\textup{Petr}}(\eta=0.5)=6.27(1+z)^{-0.92\pm0.03}$ (orange dashed line), and $R_{\textup{Petr}}(\eta=0.8)=3.10(1+z)^{-0.80\pm0.03}$ (green solid line).}
\label{fig:massrz}
\end{figure}

Comparing the median sizes of the galaxies in our sample at different epochs shows how the sizes evolve with redshift. Figure \ref{fig:massrz} shows the evolution of the median corrected radii of our mass selected sample. The blue circles show the evolution of the $R_{\textup{Petr}}(\eta=0.2)$ values, the orange diamonds show the evolution of $R_{\textup{Petr}}(\eta=0.5)$, and the green squares show the evolution of $R_{\textup{Petr}}(\eta=0.8)$. There is a clear change in each of the radii measurements from $z = 7$ to $z = 1$, particularly in the case of $\eta = 0.2$. We find that $R_{\textup{Petr}}(\eta=0.2)$ increases by a factor of 3.78 $\pm$ 0.39 from z = 7 to z = 1 whereas $R_{\textup{Petr}}(\eta=0.8)$ increases by a factor of 3.20 $\pm$ 0.19. We fit a simple power-law relation to the median measured sizes for each $\eta$ value of the form 

\begin{equation}
R_{\textup{Petr}}(\eta) = \alpha(1+z)^\beta \textup{kpc}.
\label{eq:fit}
\end{equation}

\noindent The values we find for $\alpha$ and $\beta$ for each of the methods and $\eta$ values can be seen in Table \ref{tab:fits}. The values for $\alpha$ and $\beta$ for the mass selected sample can be seen in columns 2 and 3. In Figure \ref{fig:massrz}, the power-law fits for the full samples are shown as a blue dotted line for $R_{\textup{Petr}}(\eta=0.2)$, an orange dashed line for $R_{\textup{Petr}}(\eta=0.5)$, and a green solid line for $R_{\textup{Petr}}(\eta=0.8)$. 

\begin{table}
\centering
\caption{The fits determined for both the mass selected and number density selected samples as given by equation \ref{eq:fit}.}
  \begin{tabular}{ccccc}
  \hline 
  \hline
  & \multicolumn{2}{c}{Mass} & \multicolumn{2}{c}{Number Density} \\
  $\eta$ & $\alpha$ & $\beta$ & $\alpha$ & $\beta$ \\
  \hline
  0.2 & 11.68$\pm$0.16 & -0.97$\pm$0.03 & 12.62$\pm$1.10 & -0.82$\pm$0.14 \\
  0.5 & 6.27$\pm$0.10 & -0.92$\pm$0.03 & 4.57$\pm$0.48 & -0.53$\pm$0.16 \\
  0.8 & 3.10$\pm$0.04 & -0.80$\pm$0.03 & 2.66$\pm$0.18 & -0.67$\pm$0.11 \\
  \label{tab:fits}
  \end{tabular}

\end{table}

\subsection{Number Density Selected Sample}

Instead of selecting a sample of galaxies by their mass, in this selection we create a sample of galaxies based on a constant number density. This method has been used in a number of previous studies to examine galaxy formation and evolution over a large redshift range \citep{dokkum10, papovich11, conselice13, ownsworth16}. This has been proven to have several advantages. Although the stellar mass grows through star formation and minor mergers, the number density of galaxies above a given density threshold is invariant with time in the absence of major mergers or extreme changes of star formation \citep{ownsworth16}. Selecting galaxies through this method directly tracks the progenitors and descendants of massive galaxies at all redshifts \citep[e.g.][]{mundy15, ownsworth16}. 

\begin{figure}[!ht]
\includegraphics[width=0.475\textwidth]{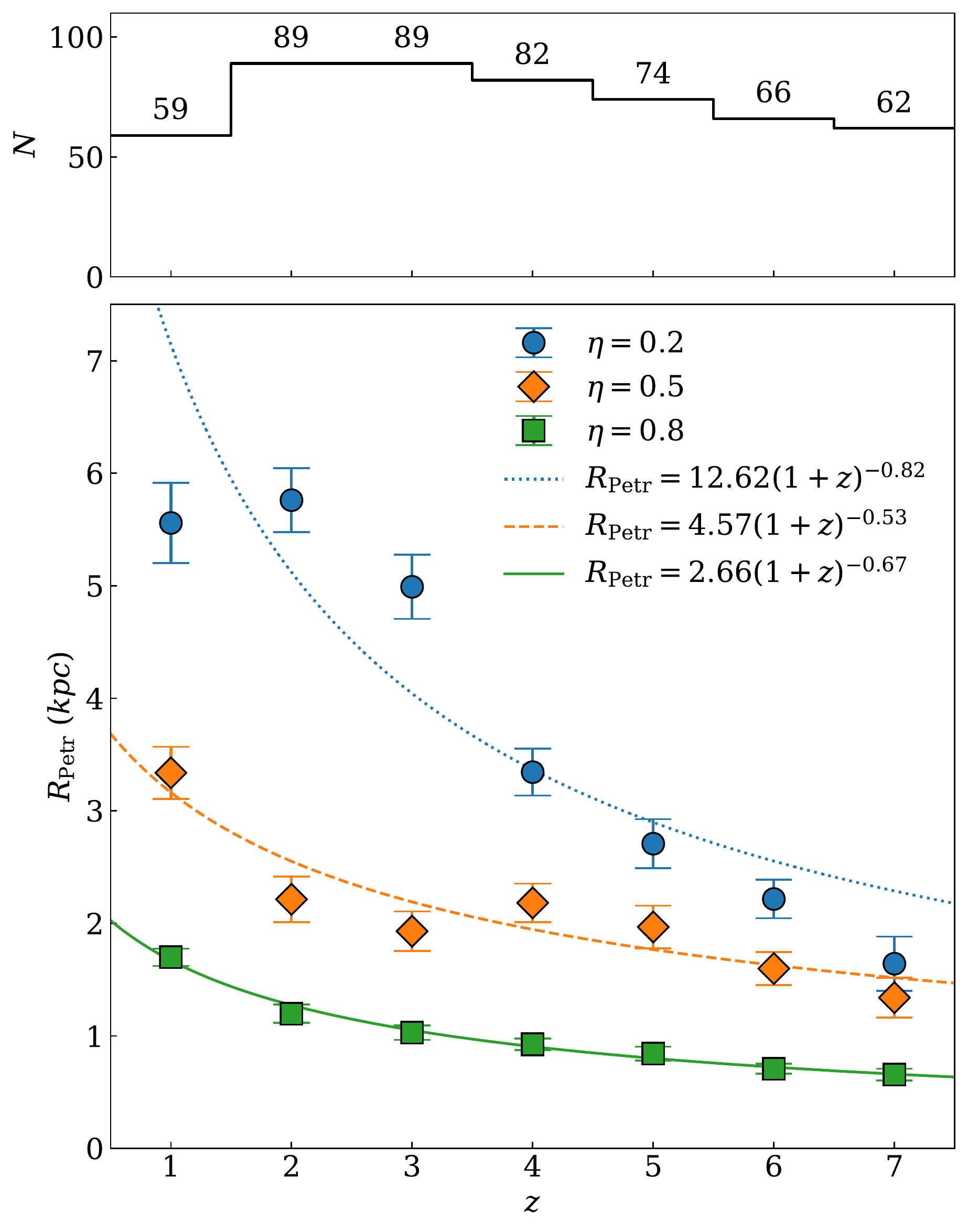}
\caption{Top: Histogram showing the distribution and number of galaxies in each redshift bin. Bottom: Evolution of the average Petrosian radius through redshift for the number density selected sample where galaxies within a constant number density of 1$\times$10$^{-4}$ Mpc$^{-3}$ are selected. Each point is the median value for each redshift bin with the error bars showing the standard error. Blue circles show how $R_{\textup{Petr}}(\eta=0.2)$ changes, orange diamonds show $R_{\textup{Petr}}(\eta=0.5)$, and green squares show $R_{\textup{Petr}}(\eta=0.8)$. A power-law relation is fit to each radius and we find $R_{\textup{Petr}}(\eta=0.2)=12.62(1+z)^{-0.82\pm0.14}$ (blue dotted line), $R_{\textup{Petr}}(\eta=0.5)=4.57(1+z)^{-0.53\pm0.16}$ (orange dashed line), and $R_{\textup{Petr}}(\eta=0.8)=2.66(1+z)^{-0.67\pm0.11}$ (green solid line).}
\label{fig:ndrz}
\end{figure}

We select a sample of galaxies using a constant number density of 1$\times$10$^{-4}$ Mpc$^{-3}$, yielding a sample size of 521 galaxies. Figure \ref{fig:ndrz} shows the evolution of the median corrected radii for this selected sample. As for Figure \ref{fig:massrz}, the blue circles represent $R_{\textup{Petr}}(\eta=0.2)$, orange diamonds represent $R_{\textup{Petr}}(\eta=0.5)$ and green squares represent $R_{\textup{Petr}}(\eta=0.8)$. We fit a power-law to the median sizes. We show these fits in Table \ref{tab:fits} and in Figure \ref{fig:ndrz} as a blue dotted line for $R_{\textup{Petr}}(\eta=0.2)$, an orange dashed line for $R_{\textup{Petr}}(\eta=0.5)$, and a green solid line for $R_{\textup{Petr}}(\eta=0.8)$. We find that $R_{\textup{Petr}}(\eta=0.2)$ changes by a factor of 3.39 $\pm$ 0.54 over the redshift range $1 < z < 7$, a much more significant change compared to a factor of 2.59 $\pm$ 0.24 for $R_{\textup{Petr}}(\eta=0.8)$ over the same redshift range.

\subsection{Inner Versus Outer Regions}

\begin{figure}[!ht]
\includegraphics[width=0.475\textwidth]{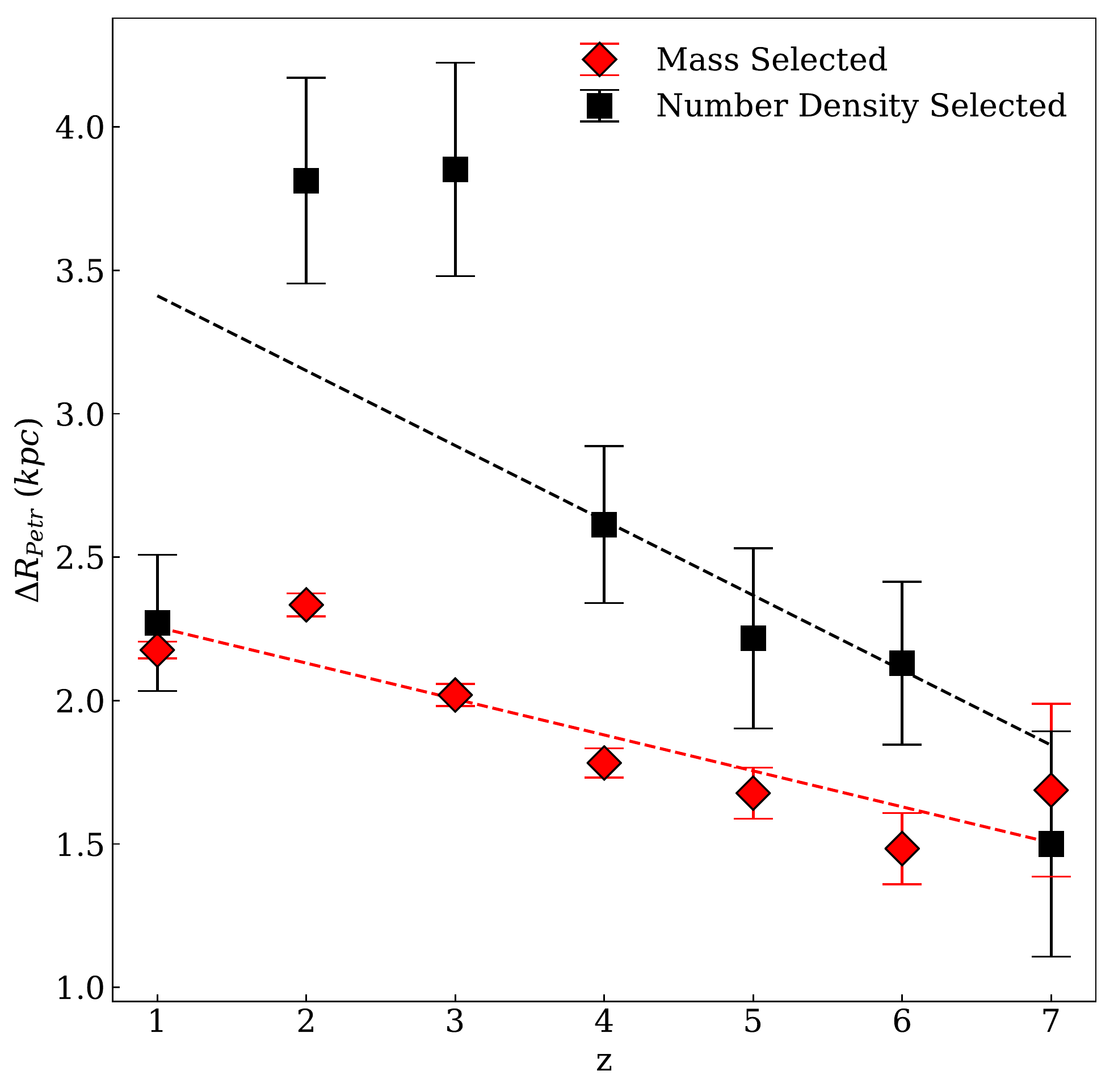}
\caption{Evolution in the normalised median difference between $R_{\textup{Petr}}(\eta=0.2)$ and $R_{\textup{Petr}}(\eta=0.8)$ for each redshift bin. This parameter, $\Delta R_{\textup{Petr}}$, is shown in Equation \ref{eq:deltar}. The mass selected sample is shown as the red diamonds and the number density selected sample is shown as the black squares. For both samples, there is an increase in $\Delta R_{\textup{Petr}}$. The number of galaxies in each redshift bin are the same as in Figure \ref{fig:massrz} for the mass selected sample and Figure \ref{fig:ndrz} for the number density selected sample.}
\label{fig:deltar}
\end{figure}

In order to determine where the radius changes the most, we plot the normalised difference between the median $R_{\textup{Petr}}(\eta=0.2)$ and $R_{\textup{Petr}}(\eta=0.8)$ against redshift in Figure \ref{fig:deltar} where $R_{\textup{Petr}}(\eta=0.2)$ corresponds to the outer edges of a galaxy and $R_{\textup{Petr}}(\eta=0.8)$ corresponds to the inner regions of a galaxy. The normalised difference in the radii is given by

\begin{equation} 
\Delta R_{\textup{Petr}} = \frac{R_{\textup{Petr}}(\eta=0.2) - R_{\textup{Petr}}(\eta=0.8)}{R_{\textup{Petr}}(\eta=0.8)}.
\label{eq:deltar}
\end{equation}

The difference between the radii measured at the outer edge and the inner regions for both the mass selected and number density selected samples is shown in Figure \ref{fig:deltar}. The mass selected sample is represented by the red diamonds and the number density selected sample by black squares. $\Delta R_{\textup{Petr}}$ increases as redshift decreases for both samples, but more significantly for the number density selected sample. 

\subsection{Galaxy Merger Sizes}

It has long been shown that galaxies increase in size as redshift decreases \citep[e.g.][]{daddi05, trujillo07, buitrago08, cimatti08, wel08, dokkum08, cassata10} but the method through which this occurs is largely unknown. We examine a sample of galaxies classified as mergers and non-mergers in order to determine whether this is a potential factor in causing the increase in size.

We identify a sample of mergers and non-mergers from the mass selected sample by utilising the CAS approach \citep{conselice03b} whereby merging galaxies are those with a high asymmetry that is larger than the clumpiness. We use the condition 

\begin{equation}
    (A > 0.35)\ \&\ (A > S)
\end{equation}

\noindent to define our sample. This method predominantly identifies only major mergers where the ratio of the stellar masses of the progenitors is at least 1:4 \citep{conselice03b, conselice06b, lotz08b}. 

\begin{figure*}[!ht]
\includegraphics[width=\textwidth]{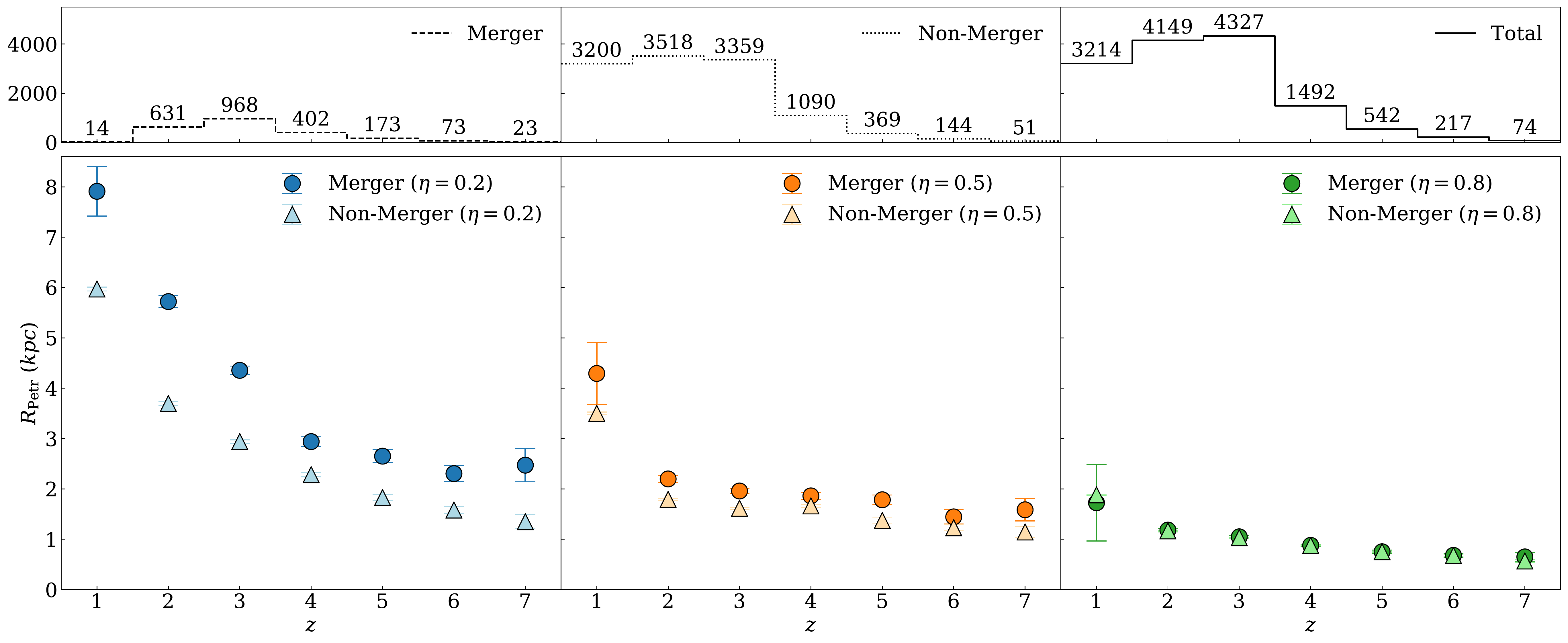}
\caption{Top: Numbers of galaxies in each redshift bin for the mergers (left), non-mergers (middle) and the total mass selected sampleh (right). Bottom: Evolution of the median Petrosian radius for mergers and non-mergers within the mass selected sample. The error bars represent the standard error, some of which are smaller than the point size. Mergers are shown by circles and non-mergers are shown by triangles. The Petrosian radius is plotted for the three different $\eta$ values with $\eta = 0.2$ shown in blue (left), $\eta = 0.5$ in orange (centre), and $\eta = 0.8$ in green (right). For all redshifts, the mergers are larger on average than the non-mergers but similar at larger $\eta$.}
\label{fig:mergers}
\end{figure*}

We show the evolution of the Petrosian radii at three different $\eta$ values in Figure \ref{fig:mergers}. Mergers are represented by circles and non-mergers are represented by triangles. The colours are the same as in Figure \ref{fig:massrz} where $R_{\textup{Petr}}(\eta=0.2)$ is shown in blue, $R_{\textup{Petr}}(\eta=0.5)$ is shown in orange, and $R_{\textup{Petr}}(\eta=0.8)$ is shown in green. For both mergers and non-mergers, the radius decreases as redshift increases irrespective of the value of $\eta$. However, the non-mergers are on average smaller than the mergers at the same redshift despite having similar masses of $10^{9.5}$M$_{\odot}$ and $10^{9.4}$M$_{\odot}$ respectively. The outermost radii change the most significantly for mergers and non-mergers, changing by factors of 3.14 $\pm$ 0.49 and 4.38 $\pm$ 0.46 respectively. $R_{\textup{Petr}}(\eta=0.8)$ changes the least with mergers and non-mergers evolving by factors of 2.62 $\pm$ 1.84 and 3.28 $\pm$ 0.18 respectively. This is a sign that there is more evolution in the outer radii sizes for mergers than for normal galaxies. During the merger process we see that galaxies are getting larger not in their centers but in their outer parts.  This is further evidence for our observational picture that galaxies are forming from the inside-out.

\section{Discussion} \label{sec:summ}

Using the GOODS North and GOODS South fields of the CANDELS data set, we present an analysis of the sizes of a sample of galaxies in the redshift range 1 $\leq$ z $\leq$ 7. We also present a new method of removing nearby unrelated field objects from images by making use of a 2-dimensional Lyman-break method at $\lambda_{rest}$ = 912{\AA}. Using these processed images, we measure the redshift-independent Petrosian radius of each galaxy at $\eta$ values of 0.2, 0.5, and 0.8 and determine how these radii change with redshift. The measured radii are then corrected to remove any effects from the PSF.  

At high redshifts i.e. $z > 3$, the size distribution of the galaxies within the full sample is dominated by galaxies with a small Petrosian radius with the median size increasing as redshift decreases. This is true for all values of $\eta$. Typically the effective radius has been used to measure the size evolution of galaxies. For example, \cite{shibuya15} find sizes that decrease significantly toward high z, no matter what statistic is used. In addition, a small sample of galaxies at $z \sim 9-10$ studied by \cite{holwerda15} were found to have mean size of $0.5\pm0.1$kpc which is consistent with extrapolated low-redshift data.

In order to remove biases introduced by the detection limits of the surveys, we select a mass-complete sample of galaxies from the main sample within the mass range $10^9$M$_{\odot} \leq M_*\leq 10^{10.5}$M$_{\odot}$. At $z = 7$, we find the average size of a galaxy is $R_{\textup{Petr}}(\eta=0.2) = 1.58 \pm 0.16$ kpc, a factor of 3.78 $\pm$ 0.39 smaller than $R_{\textup{Petr}}(\eta=0.2)$ at $z = 1$. By applying a simple power-law fit to each of the radii, we see the radii change as $(1+z)^\beta$ where $\beta < 0$ in Figure \ref{fig:massrz}. We fit $\beta = -0.97 \pm 0.03$ for $R_{\textup{Petr}}(\eta=0.2)$ and $\beta = -0.80 \pm 0.03$ for $R_{\textup{Petr}}(\eta=0.8)$ which shows that the size evolution is, on average, faster for the outer regions of the galaxies than for the inner regions. These fits are however shallower than other studies that use a simple fitted half-light radius e.g. \cite{allen17} who find $\beta = -0.89$ for a mass-complete ($M_* > 10^{10}$M$_{\odot}$) sample of galaxies from the FourStar Galaxy Evolution Survey over a redshift range of $z = 1-7$. Similarly, \cite{wel14} determine the size evolution of a sample of galaxies with $M_* \sim 10^{10}$M$_{\odot}$ to be steep with $\beta = -1.1$.

A similar result is achieved when measuring the sizes of a number density selected sample at a limit of $1 \times 10^{-4}$Mpc$^{-3}$. The median size for this selection at $z = 7$ is 1.64 $\pm$ 0.24 kpc, a number similar to that found at the same redshift for the mass selected sample. The value at $z = 1$ for the number density selected sample is a factor of 3.39 $\pm$ 0.54 larger than that at $z = 7$. The size evolution for this sample therefore evolves at a similar rate to the previous mass selected sample. 

We again find that the evolution can be fit as a power-law of the form $(1+z)^{\beta}$. For $R_{\textup{Petr}}(\eta=0.2)$, we find $\beta = -0.82 \pm 0.14$ and for $R_{\textup{Petr}}(\eta=0.8)$, we find $\beta = -0.67 \pm 0.11$. The result for the outermost radius is consistent with that of \cite{bouwens04} who find $\beta = -1.05$ and \cite{oesch10} who find $\beta = -1.12$ for their samples. In comparison \cite{patel13} select galaxies using a number density of $1.4 \times 10^{-4}$Mpc$^{-3}$ and find a value of $\beta = -1.16$ for quiescent galaxies which is slightly higher. The results found by these previous works are determined using the effective radii of galaxies.  

Independent of the selection method used, the outer radii of galaxies evolve with a steeper slope than the inner radii suggesting that the outer regions are growing more rapidly. This therefore suggests that mass is added to the outer regions in an inside-out formation mode. This difference in evolution between $R_{\textup{Petr}}(\eta=0.2)$ and $R_{\textup{Petr}}(\eta=0.8)$ is highlighted in Figure \ref{fig:deltar}. For each selection method, the value of $\Delta R_{\textup{Petr}}$ increases with time, showing that the outermost radius increases at a greater rate than the innermost radius. 

We furthermore split the mass selected sample into mergers and non-mergers based on the measured CAS values. We examine this as the merging of galaxies is a dominant method for forming distant galaxies and therefore we can determine how the size distribution changes during this process. Figure \ref{fig:mergers} shows that each of the different radii increase as redshift decreases for both mergers and non-mergers however on average, mergers are larger. As with the previous samples, $R_{\textup{Petr}}(\eta = 0.2)$ changes the most significantly with mergers changing by a factor of 3.11 $\pm$ 0.81 and non-mergers evolving in size by a factor of 3.98 $\pm$ 0.41. $R_{\textup{Petr}}(\eta = 0.8)$ changes the least with mergers and non-mergers changing by factors of 2.07 $\pm$ 0.24 and 2.31 $\pm$ 0.13 respectively. The outer radii change to a higher degree for mergers compared to the non-mergers, again suggesting a inside-out formation scenario whereby galaxy formation events increase the outer sizes more than the inner radii. 

This inside-out growth could be due to a number of factors with these results suggesting accretion of satellite galaxies being an important one \citep[e.g.][]{ferreras14, huertascompany16, buitrago17}. \cite{miller19} suggests the growth of the inner parts of galaxies is related closely to the star formation whereas the growth of the outer regions is linked to accretion and the relationship with galactic halos. By measuring $r_{20}$ and $r_{80}$ as opposed to the half-light radius which is more commonly used, they find that star forming galaxies are larger than quiescent galaxies in the inner regions ($r_{20}$) but the difference between the sizes of star forming and quiescent galaxies disappears at $r_{80}$.

 The results we find are consistent with previous work. For example, \cite{margalefbentabol16} determine the size evolution of a sample of two-component galaxies with stellar masses $M_* > 10^{10}$M$_{\odot}$ from CANDELS. They measure the circularised effective radius of each of the components and find that the outer components increase in size from $z = 3$ to $z = 1$ by a factor of 2 whereas the bulges, or inner components, remain roughly constant over the same redshift range. They conclude that this suggests inside-out formation with the bulges being in place early on in a galaxy's history. \cite{carrasco10} also find a similar result; by using observations of massive ($M_* \sim 4 \times 10^{11}$M$_{\odot}$) galaxies from the Palomar Observatory Wide-Field Infrared survey, they show that the outer regions of low-$z$ elliptical galaxies are denser than the high-$z$ compact massive galaxies by a factor of $\sim$ 2, confirming that mass is added in the outer edges. Therefore, it is now commonly seen in all studies that galaxies are growing in an inside-out fashion.

\section{Conclusions} \label{sec:conc}

The details of the processes that lead to the growth of galaxies through time are still largely unknown, but by measuring sizes using redshift-independent relative surface brightness metrics, we are able to determine where the size of a galaxy grows most rapidly, and therefore suggest how galaxies grow. In this paper, we present a new method of removing foreground objects from images of galaxies that makes use of the Lyman break at 912{\AA}. This allows us to reduce the risk of contamination from other objects when measuring the sizes and other properties of galaxies. The images we use to make our measurements are in the optical rest-frame at $\lambda \cong 4000{\AA}$ wherever possible. However, due to the limited availability of \textit{HST} bands, this is not possible for galaxies at $z > 3$ where we are forced to probe galaxy sizes in progressively bluer wavelengths down to the UV. We calculate the Petrosian radii of three different samples of galaxies selected from the CANDELS GOODS North and South fields and determine how these radii evolve with redshift from $z = 7$ to $z = 1$. We use the Petroisan radius as a proxy for size throughout. Overall, we find an increase in size from $z = 7$ to $z = 1$ with the outer radii increasing the most rapidly over this redshift range. This rapid growth in the outer edges suggests an inside-out formation process is causing the overall growth in galaxy size. 

We also determine how size evolves for a sample of mergers and non-mergers and find that mergers are, on average, larger than non-mergers at the outer radii for a given redshift. The outer radii evolve more rapidly than the inner radii, further supporting the idea that the size evolution of galaxies is caused by an inside-out formation process e.g. the accretion of satellite galaxies, mergers, and accretion of gas from the intergalactic medium. 

\section{Acknowledgements}

We thank the anonymous referee for all their very useful comments and suggestions which have lead to a greatly improved paper. This work is based on observations taken by the CANDELS Multi-Cycle Treasury Program with the NASA/ESA \textit{HST}, which is operated by the Association of Universities for Research in Astronomy, Inc., under NASA contract NAS5-26555. We thank the CANDELS team for their heroic work making their products and data available. We acknowledge funding from the Science and Technology Facilities Council (STFC). 




\bibliographystyle{yahapj}
\bibliography{sample}



\end{document}